# Interfacial Effects and Negative Capacitance State in P(VDF-TrFE) Films with BaTiO$_3$ Nanoparticles

Oleksii V. Bereznykov[1], Oleksandr S. Pylypchuk[*,1], Victor I. Styopkin[1,†], Serhii E. Ivanchenko[2], Denis O. Stetsenko[1], Eugene A. Eliseev[2], Zdravko Kutnjak[3], Vladimir N. Poroshin[1], Anna N. Morozovska[1,‡], and Nicholas V. Morozovsky[1,§]

[1]Institute of Physics, National Academy of Sciences of Ukraine, 46, pr. Nauky, 03028 Kyiv, Ukraine

[2]Frantsevich Institute for Problems in Materials Science, National Academy of Sciences of Ukraine, 3, str. Omeliana Pritsaka, 03142 Kyiv, Ukraine

[3] Jozef Stefan Institute of Slovenia, Ljubljana, Slovenia

## Abstract

We analyze the temperature dependences of the effective dielectric permittivity of P(VDF-TrFE) films with BaTiO$_3$ nanoparticles (with the average size 24 nm and dispersion 10 nm). We reveal significant deviations from the temperature dependences expected either for P(VDF-TrFE) films or for BaTiO$_3$ nanoparticles. Such dependences for BaTiO$_3$ nanoparticles would have a pronounced maxima corresponding to the average Curie temperature of the nanoparticle ensemble (located below 125°C in dependence on the particle average size and size distribution function) and a moderate frequency dispersion in the frequency range 0.1 kHz – 100 kHz. Instead of these expectations, a transition from a weak to a strong nonlinear temperature dependence of the dielectric permittivity is observed near the freezing temperature (near −50 °C) of P(VDF-TrFE). The transition is followed by a diffuse step-like change in the temperature range (0 – 40)°C, and a subsequent maximum of the dielectric permittivity in the P(VDF-TrFE) films with a lower content (~20 – 40 vol.%) of BaTiO$_3$ nanoparticles; or by a quasi-plateau of the dielectric permittivity in the P(VDF-TrFE) films with a higher content (~

---

* corresponding author, e-mail: alexander.pylypchuk@gmail.com
† corresponding author, e-mail: vstyopkin@gmail.com
‡ corresponding author, e-mail: anna.n.morozovska@gmail.com
§ corresponding author, e-mail: nicholas.v.morozovsky@gmail.com

50 – 70 vol.%) of BaTiO$_3$ nanoparticles. The frequency dispersion of the dielectric permittivity is significant in the vicinity of its maxima. The temperature-frequency shift of the permittivity region with a strong temperature dependence is positive. The temperature-frequency shift of the maxima is insignificant (or negative) for the P(VDF-TrFE) films with lower content of BaTiO$_3$ nanoparticles. Increasing the content of BaTiO$_3$ nanoparticles leads to a significant increase in the relative dielectric permittivity of the P(VDF-TrFE)-BaTiO$_3$ films compared to pure P(VDF-TrFE) films (from 8 to 50 at 25 °C). At the same time, the voltage response of the studied P(VDF-TrFE) - BaTiO$_3$ films to the frequency-modulated IR radiation flux has rather photoelectric than pyroelectric nature. A phenomenological model, which considers the screening charges at the interfaces, as well as dipole-dipole cross-interaction effects between the ferroelectric nanoparticles, is proposed to describe the temperature and frequency behavior of the effective dielectric permittivity. The negative capacitance state, which originates due to the interfacial effects, is predicted in the P(VDF-TrFE) films with a high content of BaTiO$_3$ nanoparticles.

## 1. Introduction

Thin films and colloids containing ferroelectric nanoparticles of various shapes and sizes are unique model objects for fundamental studies of interfacial and size effects, and dipole-dipole interactions in the ensemble of the nanoparticles [1, 2, 3, 4]. The practical interest in these nanomaterials is conditioned by their remarkable abilities for energy storage [5, 6, 7], piezoelectric and electrocaloric applications [8, 9].

Although traditional and innovative methods for synthesis, size, shape, and polar properties control are well developed for homogeneous [10] and core-shell type [11, 12] ferroelectric nanoparticles, there are several issues, which still pose challenges for preparation technology and reveal mysteries for underlaying physics even for small concentrations of quasi-spherical BaTiO$_3$ nanoparticles with size (5–50) nm [13, 14, 15] embedded in liquid crystal [1-3, 16] or polymer [4] environments. To the best of our knowledge, understanding of the interfacial effects and dipole-dipole cross-interactions in such systems is at the early stage.

Here we analyze the temperature dependences of the effective dielectric permittivity of the P(VDF-TrFE) films with different volume content of $BaTiO_3$ nanoparticles (with the average size 24 nm and dispersion 10 nm). We reveal significant deviations from the expected temperature dependences, which would have with a pronounced maxima corresponding to the average Curie temperature of the nanoparticle ensemble, located below 125°C in dependence on the particle average size and size distribution function, and moderate frequency dispersion in the frequency range 0.1 kHz – 100 kHz, characteristic for classical ferroelectric ceramics.
A phenomenological model, which considers the interfacial effects, and dipole-dipole cross-interactions between the ferroelectric nanoparticles, is proposed to describe the temperature dependences of the effective dielectric permittivity at different frequencies. The negative capacitance state, which originates due to the interfacial effects, is predicted in the P(VDF-TrFE) films with a high content of $BaTiO_3$ nanoparticles.

## 2. Samples Preparation and Microstructure Studies

The characterization of $BaTiO_3$ nanopowders used here is presented in Ref.[4]. The powder consists of small quasi-spherical nanoparticles, whose sizes vary from 17 nm to 47 nm, and the size distribution function is asymmetric (close to the truncated Maxwellian distribution) with the sharp maximum (dispersion 7 nm) and the particle average size is 24 nm. The phase composition of the nanopowder was determined by the Rietveld refinement of the X-ray spectrum as 7 wt.% of orthorhombic $BaCO_3$ and 93 wt.% of tetragonal $BaTiO_3$ with a relatively small tetragonality ratio, $c/a$=1.0059, (in comparison with the value 1.011 corresponding to the bulk $BaTiO_3$ single crystal [17]). The small tetragonality indicates the proximity of the size-induced paraelectric phase transition in the smallest $BaTiO_3$ nanoparticles [17], which is confirmed by the studies of ceramics spark-plasma sintered from the particles [18].

Using these BaTiO$_3$ nanoparticles, we prepared the tape-casted films and layers with a thickness ranging from 3 to 390 μm, which contain different content of the nanoparticles. Two representative compositions, which nominally contain 38.7 vol.% and 66 vol.% of BaTiO$_3$ nanoparticles, were selected for further studies; and their parameters are listed in **Table 1**. The P(VDF-TrFE) films and layers were prepared from Piezotech FC45, a copolymer of vinylidene fluoride and trifluoroethylene [Poly(VDF-co-TrFE) 55/45], supplied by Arkema Piezotech (France), which was dissolved in cyclohexanone (Lab-Scan, Ireland) using a magnetic stirrer (RTC Basic, IKA, Germany) to prepare a 10 wt.% polymer solution. Polymer tapes were fabricated by casting the solution onto silicone-coated Mylar® film (DuPont, USA).

**Table 1.** Parameters of the studied films: P(VDF-TrFE), P(VDF-TrFE) - BaTiO$_3$ nanoparticles

| Film description and thickness | BaTiO$_3$ nanopowder content, vol. % | Polymer content, vol. % | Plasticizer content, vol. % | Surfactant content, vol. % |
|---|---|---|---|---|
| P(VDF-TrFE) (120 μm) | 0 (nominal)<br>< 1 (from SEM) | 100 (nominal) | 0 | 0 |
| sample 1 (3 μm)<br>sample 2 (8 μm) | 38.7 (nominal)<br>2 - 15 (from SEM)<br>(20 ± 5) (from fitting) | 49.5 (nominal) | 8.5 | 3.2 |
| sample 3 (390 μm) | 66 (nominal)<br>(46 ± 4) (from SEM)<br>(35 ± 5) (from fitting) | 34 (nominal) | 0 | 0 |

Scanning electron microscopy (SEM) mainly uses slow secondary electrons (SE), which energy is less than 50 eV, and faster backscattered electrons (BSE), which energy is greater than 50 eV, for imaging of studied samples. The signal level of these electrons is affected by many factors, including the chemical composition of the samples. The yield of SE increases with increase in the atomic number Z of the studied material up to $Z = 20$, and then remains virtually unchanged. The yield of BSE increases continuously with increasing Z, and the dependence of the yield on Z is linear up to $Z = 30$. Therefore, BSEs are preferable to determine the chemical composition of the material, because its signal depends mainly on the average atomic number Z of the material. Due to the reason the BSE mode is referred to as Z-contrast more hereinafter.

SEM images of nominally pure P(VDF-TrFE) films and P(VDF-TrFE)-BaTiO$_3$ films are shown in **Fig. 1** and **2**, respectively. The images corresponding to the SE and Z-contrast modes confirm that all studied films are inhomogeneous. Note that the clusters of BaTiO$_3$ nanoparticles should have a significantly higher average atomic number ($Z = 33.9$) than the P(VDF-TrFE) material ($Z = 7.12$).

SEM images of the P(VDF-TrFE) film, obtained without tilting the sample, show that the film consisted of P(VDF-TrFE) globules with sizes 25-35 µm, and pores with sizes 10-25 µm, which are observed between the globules (see **Fig. 1(a)** and **1(b)**). A small amount of single bright particles with the smallest size 0.1-0.7 µm and significantly higher Z-contrast are observed at the surface of the film. These particles can be attributed to big clusters of BaTiO$_3$ nanoparticles, which probably remained due to the insufficient cleaning of the set-up for the film preparation. However, such small concentration of nanoparticles cannot influence significantly on the electrophysical properties of the films.

To study in detail the surface and morphology of pores in the P(VDF-TrFE) film, we analyzed the SE and Z-contrast images of the film tilted at the angle of 60 degrees (see **Fig. 1(c)-1(f)**). As it can be seen from **Fig. 1(c)-1(f)**, the surface of pores is self-crossed, and there is very small amount of single bright particles with the sizes of 0.1 – 0.5 µm.

Next, we studied the cut of the P(VDF-TrFE) film (see **Fig. 1(g)-1(i)**). In this case, the film cut is tilted at the angle of 60 degrees to the primary electron probe. A small amount of single bright particles with significantly larger Z are observed at the film cut. The particles are distributed both individually and in the form of clusters of sizes up to 1 µm. The minimum size of the bright particles visible in the images is about 50 nm (see e.g., **Fig. 1(i)**) Thus it can be individual $BaTiO_3$ nanoparticles, or their clusters of different sizes. Thus, the images in **Fig. 1** give us the necessary information about the microstructure of P(VDF-TrFE) films with single clusters of the nanoparticles.

(a) SE – mode of the P(VDF-TrFE)    (b) Z – contrast of the P(VDF-TrFE)

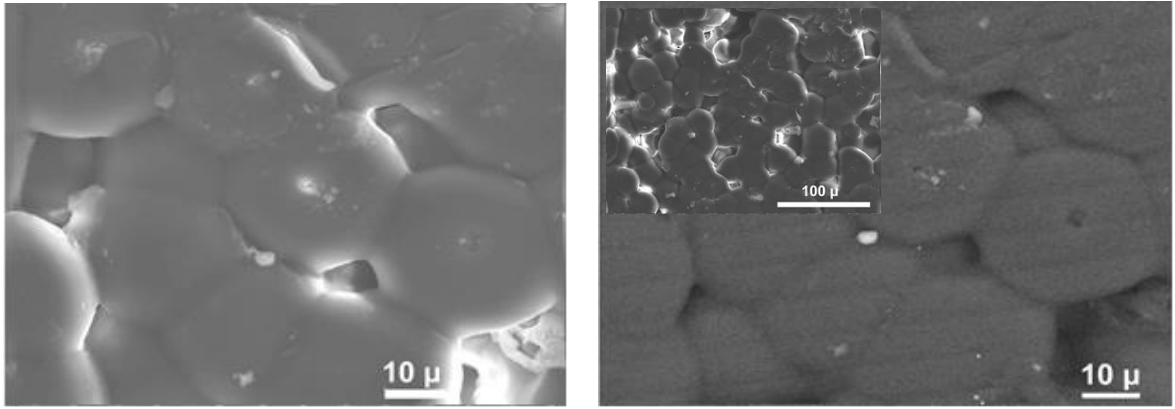

(c)-(f) SE – mode of the P(VDF-TrFE) film tilted at 60 degrees

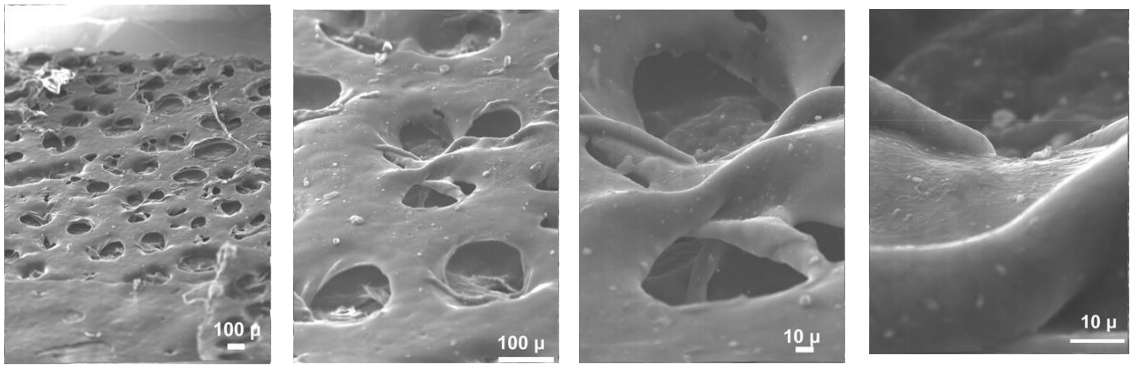

SEM imaging of the P(VDF-TrFE) film cut (tilted at 60 degrees)

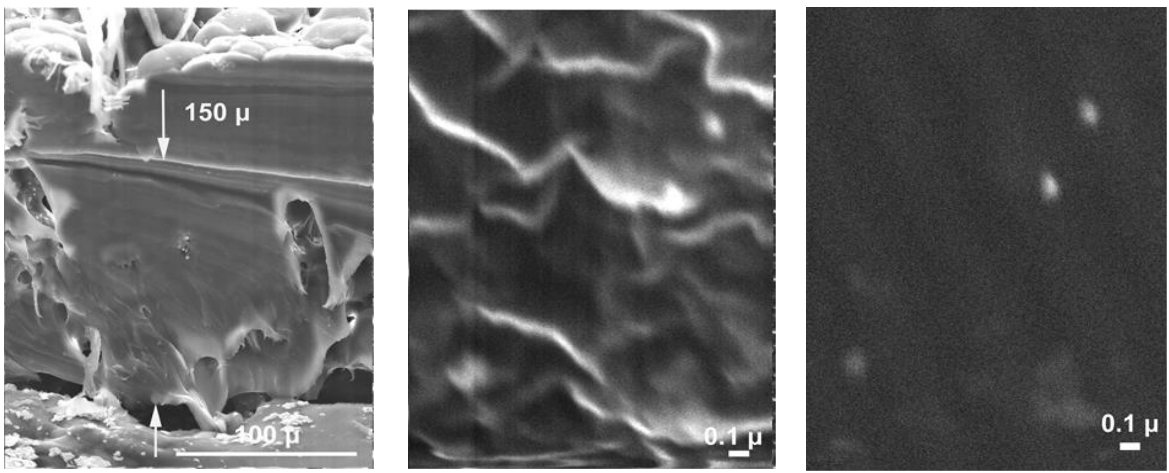

(g) SE – contrast            (h) SE – mode            (i) Z – contrast

**Figure 1.** SEM images of the P(VDF-TrFE) film. The images **(a)** and **(b)** correspond to direct incidence of the electron beam, the images **(c)-(f)** correspond to the film tilted

at 60 degrees, and the images **(g)-(i)** correspond to the film cut tilted at 60 degrees.

SEM images of samples 1 and 2, obtained without and with the sample tilt, show darker and lighter areas, which we attribute to the remainder of the P(VDF-TrFE) granules (see **Fig. 2(a)-2(d)**). The size of the inhomogeneities varies from 100 nm to several microns, which demonstrate the film inhomogeneity at the macro- and micro-scale. Bright areas with sizes of 60 - 200 nm can be attributed to the clusters of BaTiO$_3$ nanoparticles, since Ba atoms have a large atomic number Z ($Z_{Ba} = 56$). These bright areas have also inhomogeneous Z-contrast (**Fig. 2(e)** and **2(g)**). The surface itself is rather uniform, although there are areas with different brightness in Z-contrast. But the difference between the brightness of these areas is not very large.

By determining the relative level of Z-contrast signal from the SEM images, it is possible to estimate the volume concentration of BaTiO$_3$ clusters, which varies from 2 to 15 vol.% for different areas being significantly smaller than the nominal concentration of the particles content in the samples 1 and 2 (namely, 38 vol. %, see **Table 1**). We relate the significant discrepancy with the circumstance that individual nanoparticles smaller than 30 nm are below the resolution limit of the SEM, as well as recognize that undesirable clusterization of the 25-nm nanoparticles occurs in large amounts.

To study the relief of the sample 3, it was tilted at an angle of 60 degrees. The relief of the sample is inhomogeneous (**Fig. 2(g) - 2(i)**). The sizes of the relief features vary from 100 nm to 5 - 10 µm; and the particles with different brightness and sizes ranging from 200 nm to 2-3 µm are observed. The particles with increased Z-contrast have sizes from 100 nm to 2.5 µm. To estimate the content of BaTiO$_3$ nanoparticles, a reference Si plate ($Z_{Si} = 56$) was placed at a distance of 0.5 mm from the sample 3. By comparing the integral signals from the Si plate and the sample 3, the content of BaTiO$_3$ was estimated as $(46 \pm 4)$ vol %. This value is significantly lower than the nominal content of BaTiO$_3$ nanoparticles (70 vol.%), and their sizes are significantly larger than the average size 25 nm.

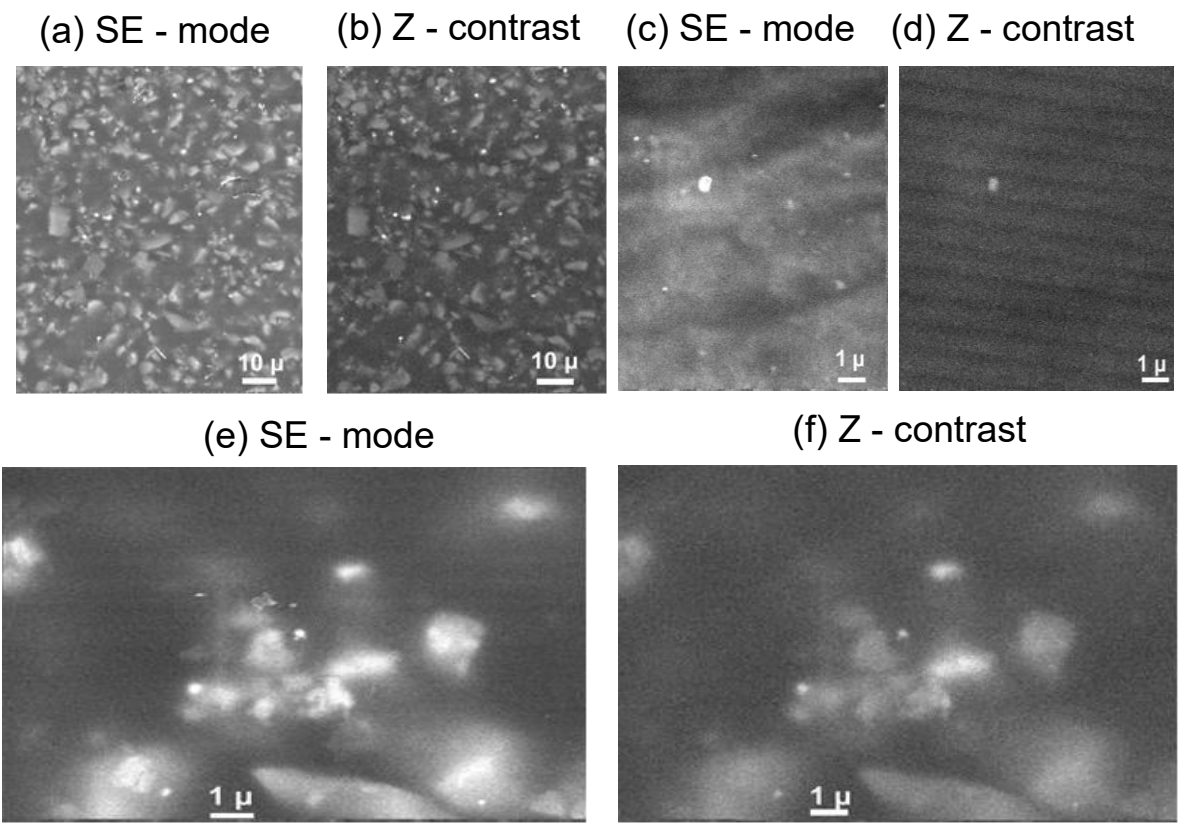

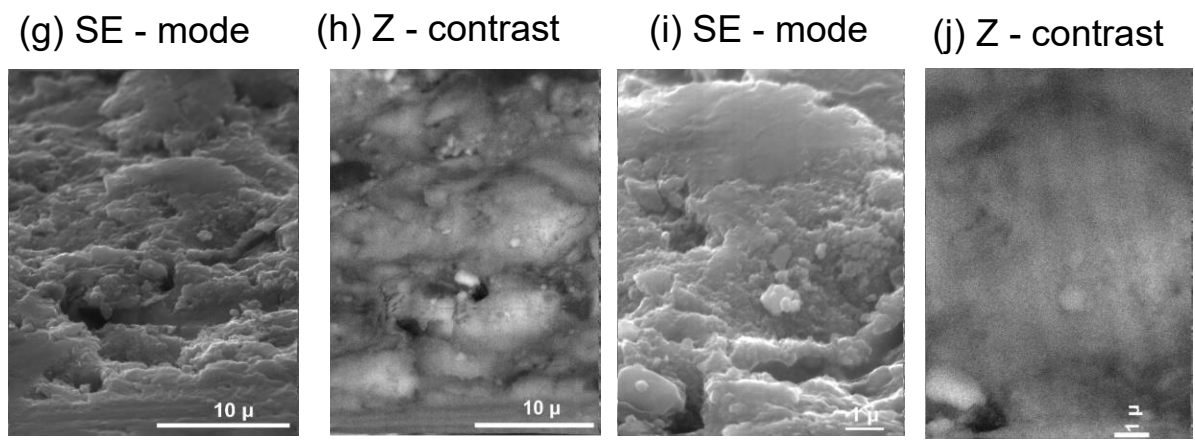

**Figure 2.** SEM images of the free-standing P(VDF-TrFE) film with BaTiO$_3$

nanoparticles (samples 1, 2 and 3). The images **(e)** and **(f)** correspond to direct incidence of the electron beam, other images correspond to the film tilted at 60 degrees.

## 3. Measurement Methodic

The P(VDF-TrFE) films with nanoparticles prepared for dielectric measurements have the form of thin films and layers deposited on a metallized textolite substrate, which acts as a bottom electrical contact. The top contacts were made by deposition of silver electrodes (see **Fig. 3(a)**). We also studied P(VDF-TrFE) films with silver electrodes deposited symmetrically on their sides (see **Fig. 3(b)**). After electrodes deposition the samples were placed between two foiled fiberglass sheets, which are the top and bottom contacts, respectively. Indium contacts were additionally used to improve the conductivity between the plates and the samples.

Measurements of the capacitance and dielectric losses of the P(VDF-TrFE) – $BaTiO_3$ films were carried out for frequencies from 0.1 kHz to 100 kHz with an LCR-meter UNI-T UT612 in the temperature range from -200°C to 100°C using a cryostat and a heating chamber with a temperature control (a germanium thermistor and thermocouple-type thermometer, respectively) (see **Fig. 3(c)**). The accuracy of the temperature control was about 0.1°C, and the capacitance measurement error was less than 0.5%.

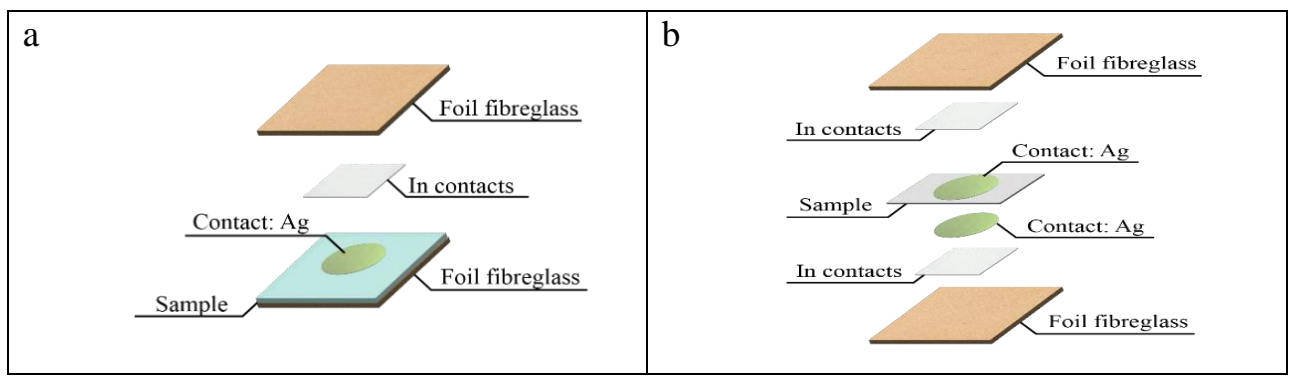

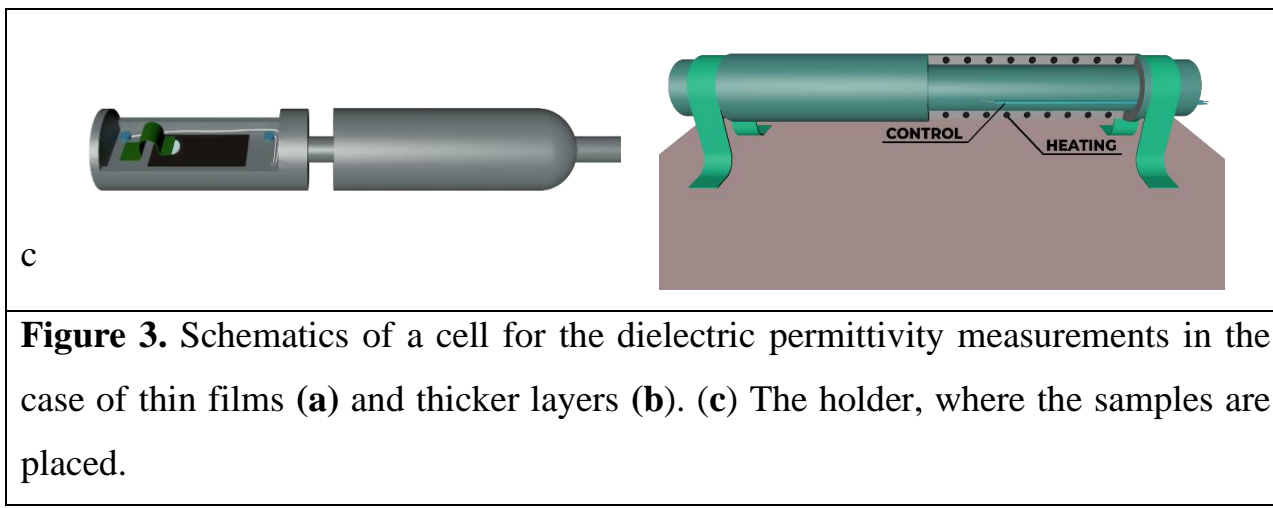

**Figure 3.** Schematics of a cell for the dielectric permittivity measurements in the case of thin films **(a)** and thicker layers **(b)**. **(c)** The holder, where the samples are placed.

## 4. Experimental Results

Below we analyze temperature dependences of the effective dielectric permittivity $\varepsilon(T)$ obtained for the studied films and layers at different frequencies. **Figures 4(b)-(c)** show the changes in the $\varepsilon(T)$ dependences with increasing the content of nanoparticles. **Figure 4(a)** shows $\varepsilon(T)$ for the pure P(VDF-TrFE) film. **Figures 4(b), 4(c)** and **4(d)** show the $\varepsilon(T)$ dependences for the P(VDF-TrFE) - $BaTiO_3$ samples 1, 2 and 3, respectively (see **Table 1**).

In the temperature range from -196°C to 150°C, the dielectric permittivity value varies from 4 to 16 for pure P(VDF-TrFE) film (**Fig. 4(a)**); from 7 to 29 for the sample 1 (**Fig. 4(b)**), from 14 to 33 for the sample 2 (**Fig. 4(c)**), and from 37 to 100 for the sample 3 (**Fig. 4(d)**). The unusual variation of $\varepsilon(T)$ evidences the dipole-type interactions of the $BaTiO_3$ nanoparticles with the P(VDF-TrFE) matrix on the one hand. On the other hand, it demonstrates the existence of a direct contribution of $BaTiO_3$ nanoparticles to the dielectric response of P(VDF-TrFE) films due to the interfacial effects in the effective medium.

At low temperatures, the temperature dependence of ε(T) is weak for all P(VDF-TrFE) and P(VDF-TrFE)-$BaTiO_3$ films. For pure P(VDF-TrFE) film, a weak trend towards saturation is observed up to −100 °C (see **Fig. 4(a)**). The ε(T) dependences obtained for pure P(VDF-TrFE) films are similar to those known for the melt-crystallized PVDF films [19], for protonated and deuterated PVDF films [20], as well as for PVDF-PZT composites [21] and P(VDF-TrFE) copolymers [22]. For all types of PVDF layers, the increase in ε(T) begins in the vicinity of $T_f \approx 223$ K (–50 °C), where $T_f$ is the so-called "freezing" temperature of the polar system in the P(VDF-TrFE), where the maximum of dielectric losses is observed [19-21]. Sometimes this characteristic temperature is called the glass transition temperature and is denoted as $T_g$, which indicates the transition of the polar system to a glassy state, which is characterized by a certain level of disorder [22]. The step-like change in ε(T) near –50 °C is a common feature of the studied P(VDF-TrFE) and P(VDF-TrFE) - $BaTiO_3$ films.

The characteristic features of the samples 1 and 2 with a low content of $BaTiO_3$ particles are the transition from the low-temperature region with a weak dependence of ε(T) to the region of diffuse step-like change, and then to the diffuse maximum and minimum, which precedes the region of further increase in ε(T) at elevated temperatures (see **Fig. 4(b,c)**). The characteristic features of the sample 3 with a high content of $BaTiO_3$ particles are the transition from the low-temperature region with a weak dependence of ε(T) to the region of diffuse step-like change, and then to the quasi-plateau of ε(T), which precedes the region of further increase of ε(T) at elevated temperatures (see **Fig. 4(d)**). Within the region of step-like change of ε(T), the permittivity changes more strongly in the P(VDF-TrFE)-$BaTiO_3$ films than in the P(VDF-TrFE) films. The limits of this change are maximal for the sample 3 with the maximum content of $BaTiO_3$ nanoparticles.

Thus, the addition of $BaTiO_3$ nanoparticles leads to the appearance of the maximum of ε(T) at the end of the step-like change in the ε(T) for P(VDF-TrFE)-films with a lower content of the nanoparticles (see **Fig. 4(b,c)**). The addition of a high content of $BaTiO_3$ nanoparticles leads to the extension of the quasi-plateau in the ε(T) dependences towards high temperatures (**Fig. 4(d)**). The result of further heating is the maximum, that should be observed at temperatures above 100°C, but such heating would lead to the destruction of the P(VDF-TrFE) film.

The nonlinear increase in the $\varepsilon(T)$ for P(VDF-TrFE) - $BaTiO_3$ films begins at lower temperatures than for the P(VDF-TrFE) films. A significant increase in the $\varepsilon(T)$ begins near -50°C and continues up to (10 - 30)°C (depending on the frequency). In this temperature range, the value of $\varepsilon(T)$ changes more sharply in the P(VDF-TrFE) - $BaTiO_3$ films, than that of the P(VDF-TrFE) films, which leads to 2-5 times higher values of $\varepsilon(T)$ maxima. For the studied P(VDF-TrFE) - $BaTiO_3$ films, a noticeable inflection is observed in the temperature range below 0°C. As can be seen from the comparison of these figures, the values of $\varepsilon(T)$ and their temperature changes are different. We ascribe the difference to the interfacial effects and dipole-dipole interactions between $BaTiO_3$ nanoparticles through the P(VDF-TrFE) spacers.

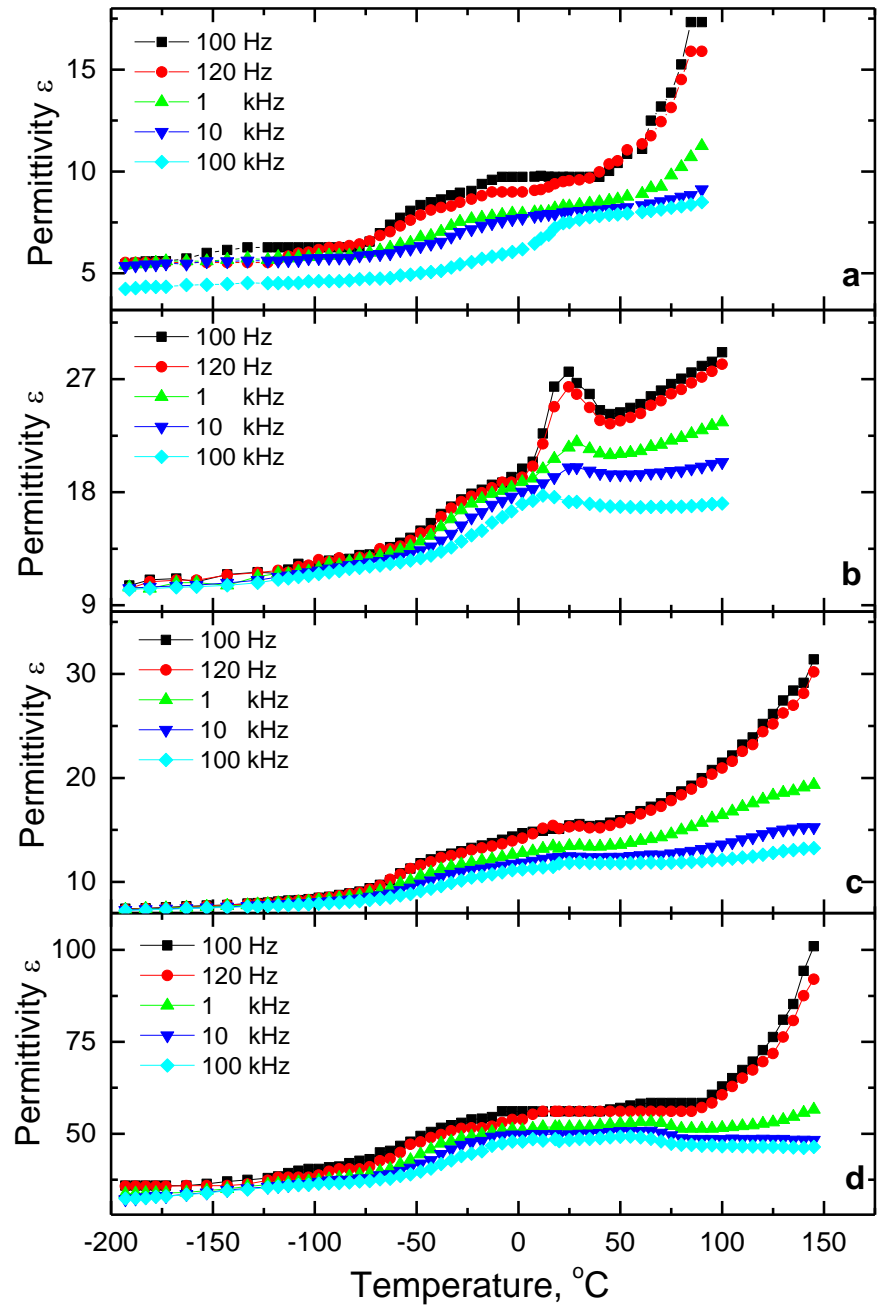

**Figure 4.** Temperature dependences of the effective dielectric permittivity at different frequencies for the pure P(VDF-TrFE) film **(a)** and P(VDF-TrFE)-BaTiO$_3$ films **(b)**-**(d)**. The dependences shown in the plots **(b)**, **(c)** and **(d)** correspond to the samples 1, 2 and 3, respectively.

**Figure 5** shows the changes in the dependences of the dielectric loss tangent tanδ(T) with increasing the content of BaTiO$_3$ nanoparticles. **Figure 5(a)** shows the dependences for the pure P(VDF-TrFE) film, **Figs. 5(b), 5(c)** and **5(d)** show the dependences for the P(VDF-TrFE)-BaTiO$_3$ samples 1, 2 and 3, respectively. For the studied P(VDF-TrFE) and P(VDF-TrFE)-BaTiO$_3$ films, there is a coincidence of the temperature limits of the features of ε(T) and tanδ(T) (compare **Figs. 4** and **Fig. 5**).

For the P(VDF-TrFE) films, the temperature positions of the diffuse step-like change of ε(T) and of the diffuse maximum of tanδ(T) practically coincides in the temperature range from –50°C to 0°C for low frequencies (~0.1 – 1 kHz) and in the vicinity of 0 °C for higher frequencies (~10 – 100 kHz) (see **Fig. 4(a)** and **Fig. 5(a)**).

For the P(VDF-TrFE) films with a lower content of BaTiO$_3$ nanoparticles, the physical picture is more complicated. For low frequencies (~0.1 kHz), the temperature positions of the diffuse maximum of ε(T) and the maximum of tanδ(T) also practically coincide in the temperature range 0 – 50°C, but for higher frequencies (~10 – 100 kHz) there is another maximum of tanδ(T) in the temperature range from –50°C to 0°C, where the diffuse step-like change of ε(T) is observed (see **Fig. 4(b,c)** and **Fig. 5(b,c)**).

For the P(VDF-TrFE) films with a high content of BaTiO$_3$ nanoparticles, the temperature position (≈ 25 °C) of the tanδ(T) maximum practically coincides with the middle of the interval (0 - 50) °C of the ε(T) quasi-plateau (see **Fig. 4(d)** and **Fig. 5(d)**).

It is also worth noting that the average temperatures of the considered diffuse features of ε(T) (namely, the average position of step-like change at −25°C), ε(T) and tanδ(T) (namely, the maxima at +25°C) for the studied samples 1, 2 and 3 are close to the temperatures −22°C and +27°C. These temperatures are characteristic for the (54/46) P(VDF-TrFE) copolymer, indicating the transition to the glass state in the PVDF and TrFE, respectively (see e.g., Fig. 11 and Fig. 13 in Ref.[23]). Following Ref. [23], the temperature difference characteristic for P(VDF-TrFE) copolymers with different VDF/TrFE ratios (from 30/70 to 90/10), we can conclude that a similar temperature difference for the P(VDF-TrFE) films with a lower content of BaTiO$_3$ nanoparticles may be related not only to the VDF/TrFE ratio, but also to the difference in the local degree of their porosity, interfacial effects, and dipole-dipole interactions of the BaTiO$_3$ nanoparticles.

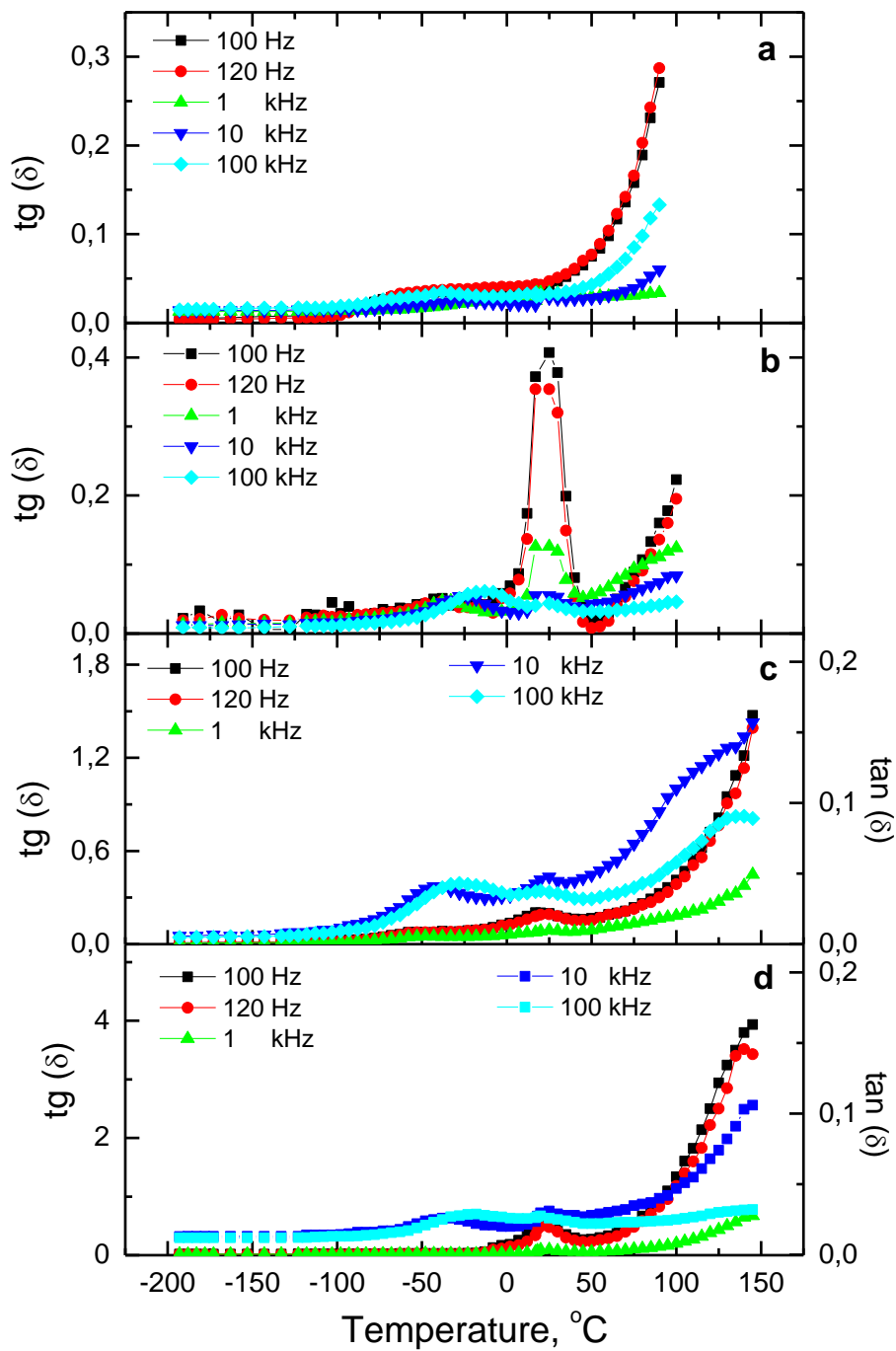

**Figure 5.** Temperature dependences of dielectric loss tangent at different frequencies for the pure P(VDF-TrFE) film **(a)** and P(VDF-TrFE)-BaTiO$_3$ films **(b)-(d)**. The dependences shown in the plots **(b)**, **(c)** and **(d)** correspond to the samples 1, 2 and 3, respectively.

For the studied porous P(VDF-TrFE) and P(VDF-TrFE)-BaTiO$_3$ films (see **Fig. 1** and **Fig. 2**), peculiarities of temperature changes in the dielectric permittivity and loss tangent (see **Fig. 4** and **Fig. 5**) could be a consequence of the interaction between the surface polar groups CH$_2$ and CF$_2$ of P(VDF-TrFE) molecular chains (namely, (-CH$_2$-CF$_2$-) for VDF and (-CHF-CF$_2$-) for TrFE) and the environment in the pores for pure P(VDF-TrFE), as well as a consequence of the additional interaction of these groups with the surface dangling bonds of BaTiO$_3$ nanoparticles placed in the P(VDF-TrFE) matrix. In the latter case, a dipole-dipole interaction occurs between the high-frequency dielectric response due to the movements of the polar groups (CH$_2$, CF$_2$ and CHF) in the chains of the P(VDF-TrFE) copolymer [(-CH$_2$-CF$_2$-) in VDF and (-CHF-CF$_2$-) in TrFE] and the low-frequency response due to the movement of the charge carriers with the participation of Ba-O- and Ti-O- bonds at the surface of BaTiO$_3$ nanoparticles. A significant increase in the BaTiO$_3$ content leads to a decrease in the thickness of the P(VDF-TrFE) layers, where the movements of polar chains become difficult due to size effects (the so-called "concentration transition" occurs [23]). For P(VDF-TrFE) films with a high content BaTiO$_3$ nanoparticles, this effect results in the suppression of the polar reaction inherent to the P(VDF-TrFE) films against the background of a general increase in the dielectric permittivity (compare **Fig. 4(b,c)** and **4(e)**, **Fig. 5(b,c)** and **5(e)**).

In the **Figure 6,** the charge-voltage (Q-V) and the current-voltage (I-V) loops of the P(VDF-TrFE) - BaTiO$_3$ films are shown. The loops were obtained by using a modified Sawyer–Tower circuit being fed by a sinusoidal 50 Hz driving voltage from a functional generator. The reference capacitor for Q-V-loops and the reference resistor for I-V-loops were used. The loops were observed "in vivo" on the screen of two-beam oscilloscope operating in X-Y-mode.

From the one side, BaTiO$_3$ grains can be surrounded by a wide shell of P(VDF-TrFE) in the P(VDF-TrFE) - BaTiO$_3$ films (see the SEM images). On the other side, the relatively low permittivity of P(VDF-TrFE) ~10 [24] (see also **Fig. 4(a)**) results in the effective dielectric decoupling of BaTiO$_3$ nanoparticles from an external electric field that is concentrated in the P(VDF-TrFE) matrix. On the other hand, the Curie temperature of BaTiO$_3$ nanoparticles, T$_C$, decreases with decreasing in the particle size $d$ and T$_C$ << 250 K for $d$ < 100 nm (see Fig. 1 in Ref. [25]). Thus, following Ref. [25], BaTiO$_3$ nanoparticles with the size 17 nm < $d$ < 47 nm, which were used in the studied P(VDF-TrFE) - BaTiO$_3$ films, are likely not ferroelectric at room temperatures.

Note that the coercive field of P(VDF-TrFE) is about 1 MV/cm, which is much higher than the electric field applied in our experiments. The voltage of 400 V was applied to the 390 μm thick layer of P(VDF-TrFE) - BaTiO$_3$ film, which corresponds to the averaged electric field of about 10.3 kV/cm (see **Figs. 6(a)** and **6(b)**), being considerably higher than the coercive field of BaTiO$_3$ single crystals (about 1 kV/cm) (see Fig. 4.9 in Ref. [26]). Thus, the applied voltage in 10 times higher the coercive one for BaTiO$_3$ single crystals cannot reverse the polarization in the studied P(VDF-TrFE) - BaTiO$_3$ film due to the field drop in the P(VDF-TrFE) matrix. The weak feature at the Q-V loop, seen in **Fig. 6(a)** around zero voltage, is associated with the non-linearity of the driving voltage. The visible high-voltage sub-linearity of the I-V loop, shown in **Fig. 6(b)**, seems to be related with the space charge effects at the BaTiO$_3$ - P(VDF-TrFE) interfaces, or at the film-electrode boundaries.

As one can see from **Figs. 6(c)** and **6(d)**, the voltage of about 20 V is enough to open the ferroelectric hysteresis loop in the 300 μm thick BaTiO$_3$ single-crystal, which correspond to the averaged coercive field about 0.7 kV/cm.

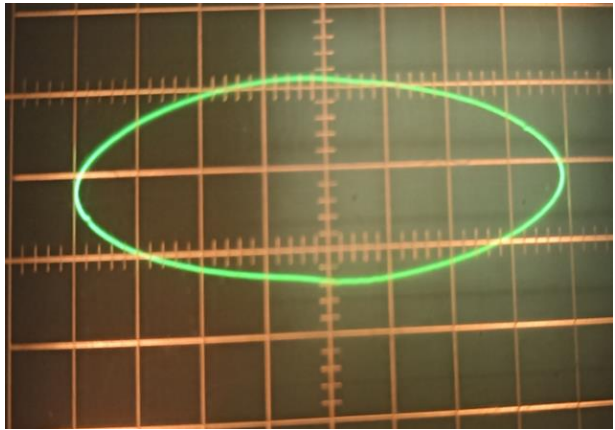 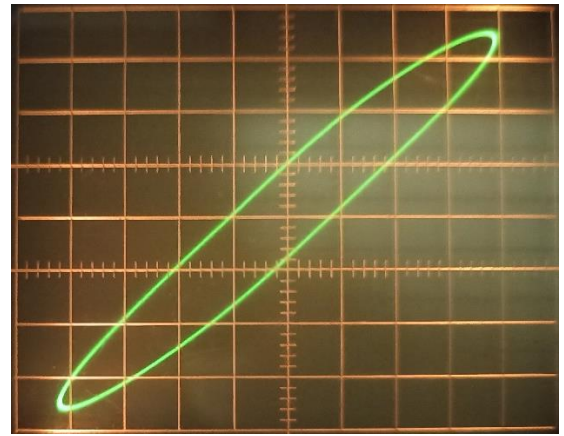

(a) (b)

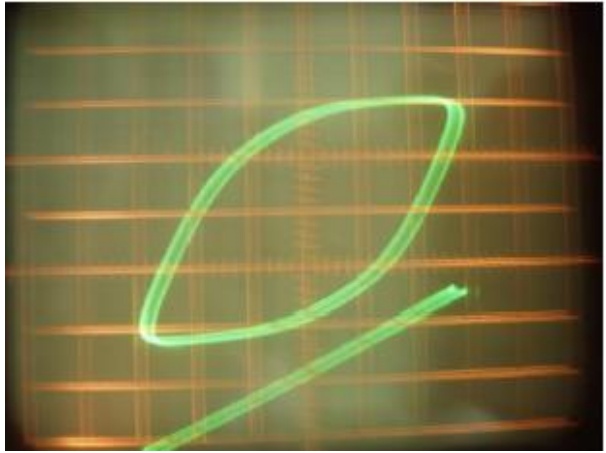 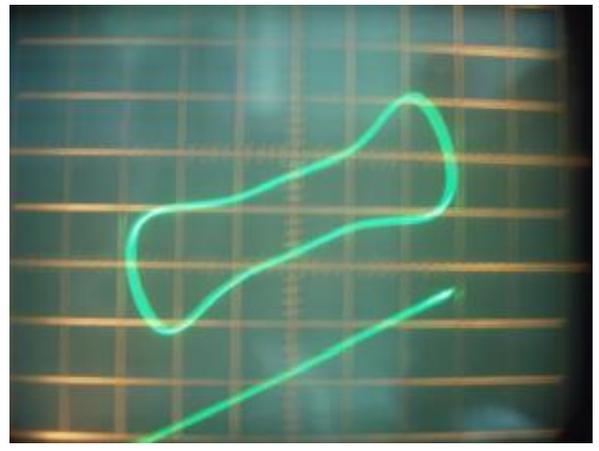

(c) (d)

**Figure 6.** Typical charge-voltage **(a)** and the current-voltage **(b)** characteristics of the P(VDF-TrFE)-BaTiO$_3$ films. The horizontal scale is 100 V/div and the vertical scale is 0.2 V/div, the reference capacitance is 0.01 µF (for the plot **(a)**) and the reference resistance is 100 kΩ (for the plot **(b)**). The charge-voltage **(c)** and current-voltage **(d)** ferroelectric polarization reversal loops of the BaTiO$_3$ single crystal plate (about 0.3 mm thick) at 50 Hz sine driving voltage. The horizontal scale is 10 V/div and the vertical scale is 0.02 V/div, the reference capacitance is 1 µF for (a) and the vertical scale is 0,1V/div, the reference resistance is 10 kΩ for (b).

As the next step, P(VDF-TrFE)-BaTiO$_3$ films were irradiated by the IR radiation flux modulated with the frequency $f_m$ from the LED driven by the sine voltage generator. The open circuit conditions were used in the sample studies. The equipment for pyroelectric measurements is described in Ref. [27].

The dependences of the electric response $U_r$ on the frequency of IR irradiation incident at both sides of the P(VDF-TrFE) - BaTiO$_3$ film are shown in **Fig. 7(a)**. For the case the values of open circuit response voltage $U_r$ are different, but the shape of obtained dependences $U_r(f_m)$ and $U_r(f_m) f_m$ is similar (compare the left and right side in **Fig. 7(a)**). For both sample sides, the values of phase shift between the IR irradiation flux and the response voltage $U_r(f_m)$ are slightly different ($\approx 10°$), but the phase $\varphi_r(f_m)$ is constant in average. The main features of the response are weak frequency dependences of $U_r(f_m)$ and close to zero $\varphi_r(f_m)$ values, which are not characteristic for the pyroelectric response of polar dielectrics in the open circuit mode.

For the open circuit conditions the response of a uniformly poled sample made from a pyroelectric material is $U_r = U_\pi(f_m) \sim (\gamma/c_1\varepsilon)/f_m$, where $\gamma$ is the pyroelectric coefficient, $c_1$ is the volume heat capacity, $\varepsilon$ is the dielectric permittivity, $f_m$ is the modulation frequency [28]. As an example of such pyroelectric response, the dependences of the open circuit pyroelectric voltage $U_\pi(f_m)$ and its phase shift $\varphi_\pi(f_m)$ via the frequency of the IR irradiation flux are presented in **Fig. 7(b)**. These dependences are obtained for commercial P(VDF-TrFE) film ("Piezotech", France). For this film, the behaviour of the $U_\pi(f_m)$ and the $\varphi_\pi(f_m)$ in open circuit conditions is typical for the uniformly poled state, when $U_\pi(f_m) \sim 1/f_m$, and so, $U_\pi(f_m)f_m$ is constant, as well as the phase $\varphi_\pi$ is a frequency independent constant equal to 90 ° (or 270°) depending on the polarization direction under the probing electrode.

PVDF-BTO film:
8x8 mm, 830 μm thick, ϕ6 mm Ag-electrodes
C = 14,4 pF, G = 0,001 μS
Open circuit electric reponse on modulated
IR-radiation flux (950 nm, 3 mW/mm$^2$)
and response-flux phase shift
(measurements without poling)

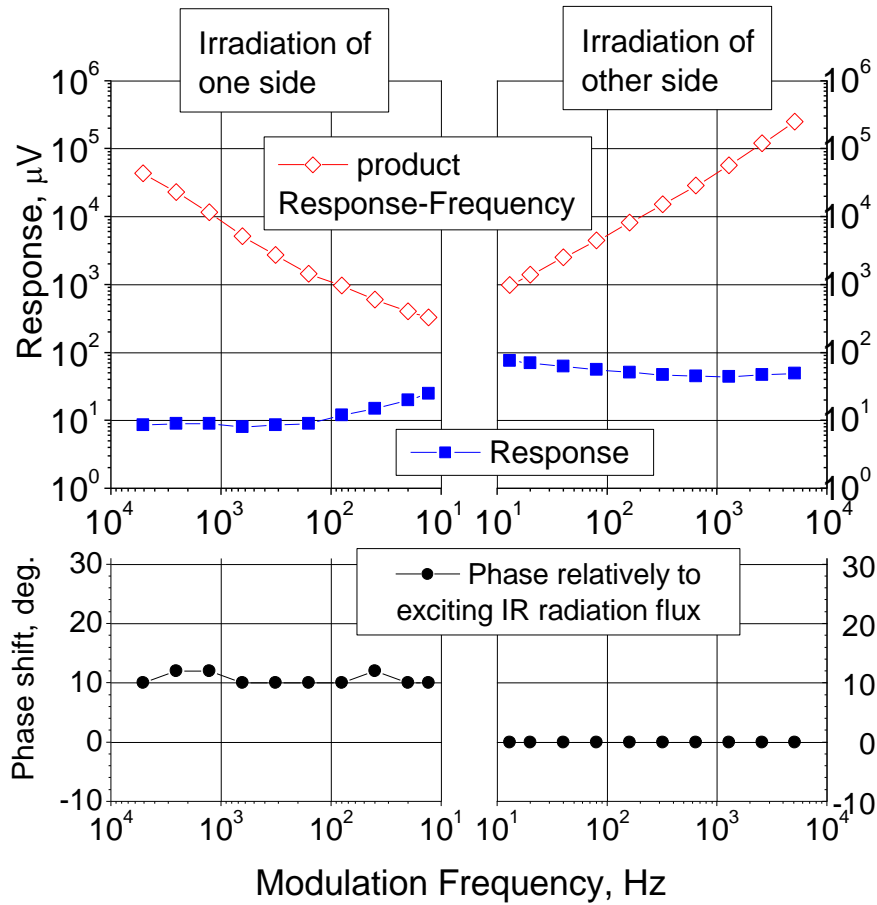

For open circuit pyroelectric response of uniformly poled sample
must be $U_\pi(f_m) \sim 1/f_m$; $U_\pi f_m \sim \gamma/c_1\varepsilon$, $\varphi_\pi(f_m) = (+/-)90° = $ const
($\gamma$ is the pyroelectric coefficient, $c_1$ is the volume heat capacity,
$\varepsilon$ is the dielectric permittivity, $f_m$ is the modulation frequency )

(a)

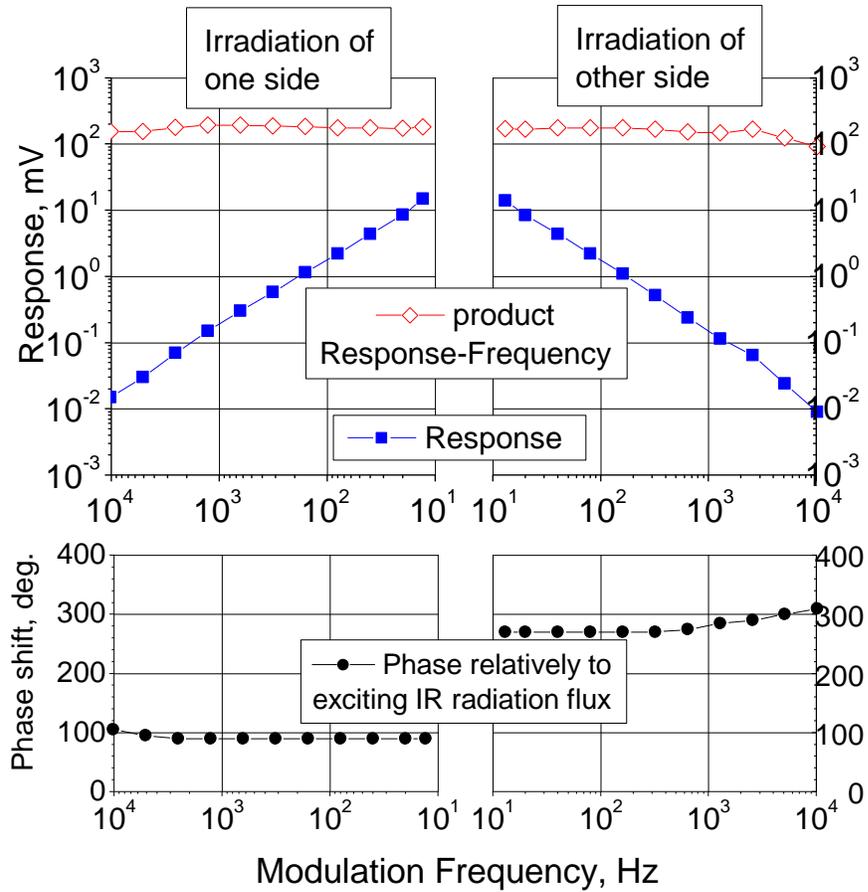

PVDF/TrFE (75/25) film
Commercial "Piezotech" (France)
Streth manufactured, Corona poled, Electroded
2,5 mm×5 mm, 14 μm thick sheet
C= 100 pF, G=0,01 μS; $\varepsilon_e$ = 12,4
Open circuit electric reponse on modulated
IR-radiation flux (950 nm, 1 mW/mm$^2$)
and response-flux phase shift

For open circuit pyroelectric response of uniformly poled sample must be $U_\pi(f_m) \sim 1/f_m$; $U_\pi f_m \sim \gamma/c_1\varepsilon$, $\varphi_\pi(f_m) = (+/-)90^o$ = const ($\gamma$ is the pyroelectric coefficient, $c_1$ is the volume heat capacity, $\varepsilon$ is the dielectric permittivity, $f_m$ is the modulation frequency )

**(b)**

**Figure 7**. Dependences of the open circuit electric response on the modulation frequency of IR-irradiation of the both sides of P(VDF-TrFE)- $BaTiO_3$ film **(a)** and the pyroelectric response of the commercial P(VDF-TrFE) film ("Piezotech", France) **(b)**.

For an in-homogeneously polarized polyethylene thin film, under probing of the both sides, the values of $U_\pi$ and the shape of obtained dependences $U_\pi(f_m)$ and $\varphi_\pi(f_m)$ are significantly different (compare the left and right panels in **Fig. 8(a)**). At low $f_m$, the $\varphi_\pi$ difference is about 180 ° that corresponds to the opposite polarization direction under both probed surfaces. Increase of $f_m$ leads to increase in $\varphi_\pi$, which can be considered as the consequence of polar and/or thermal inhomogeneity. At high $f_m$, the $U_\pi(f_m)$ dependences tends to $U_\pi(f_m) \sim 1/f_m$ and $\varphi_\pi(f_m)$ dependences tends to $\varphi_\pi(f_m) \approx$ const, which reflects the uniformity of the subsurface distribution of $\gamma/c_1\varepsilon$ value. After diaphragming the sample, so that the LED radiation fell only on the electrode, and applying a light-absorbing coating to the electrodes, the response was practically zero.

The dependences $U_r(f_m)$ and $\varphi_r(f_m)$ like those of studied P(VDF-TrFE) - $BaTiO_3$ film (**Fig. 7(a)**), were obtained for Ge diode operating also in the open circuit mode (**Fig. 8(b)**). Therefore, the voltage response of the studied P(VDF-TrFE) - $BaTiO_3$ film on the frequency-modulated IR radiation flux has rather photoelectric than pyroelectric nature.

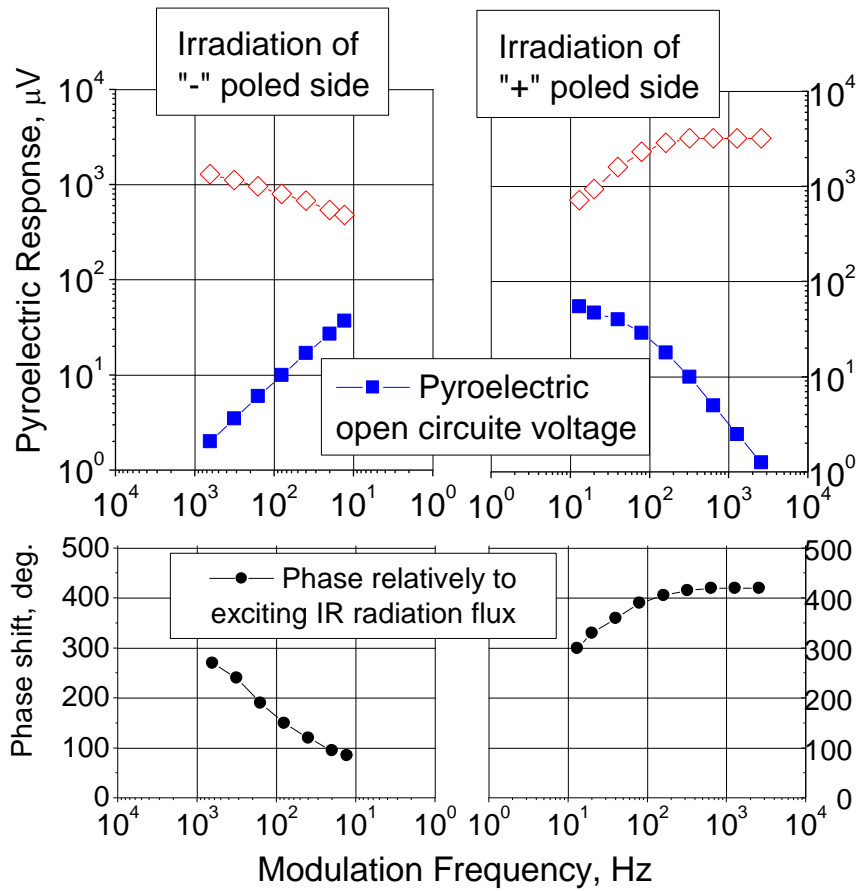

Polyethylene sheet:
6x4 mm, 70-100 μm thick,
Ag-paste electrode 4x4 mm
C = 3,3 pF, G = 0,002 μS
Pyroelectric open circuit response on modulated
IR-flux and response-flux phase shift
(first measurements)

For open circuit pyroelectric response of uniformly poled sample must be $U_\pi(f_m) \sim 1/f_m$; $U_\pi f_m \sim \gamma/c_1\varepsilon$, $\varphi_\pi(f_m) = (+/-)90° =$ const
($\gamma$ is the pyroelectric coefficient, $c_1$ is the volume heat capacity, $\varepsilon$ is the dielectric permittivity, $f_m$ is the modulation frequency )

**(a)**

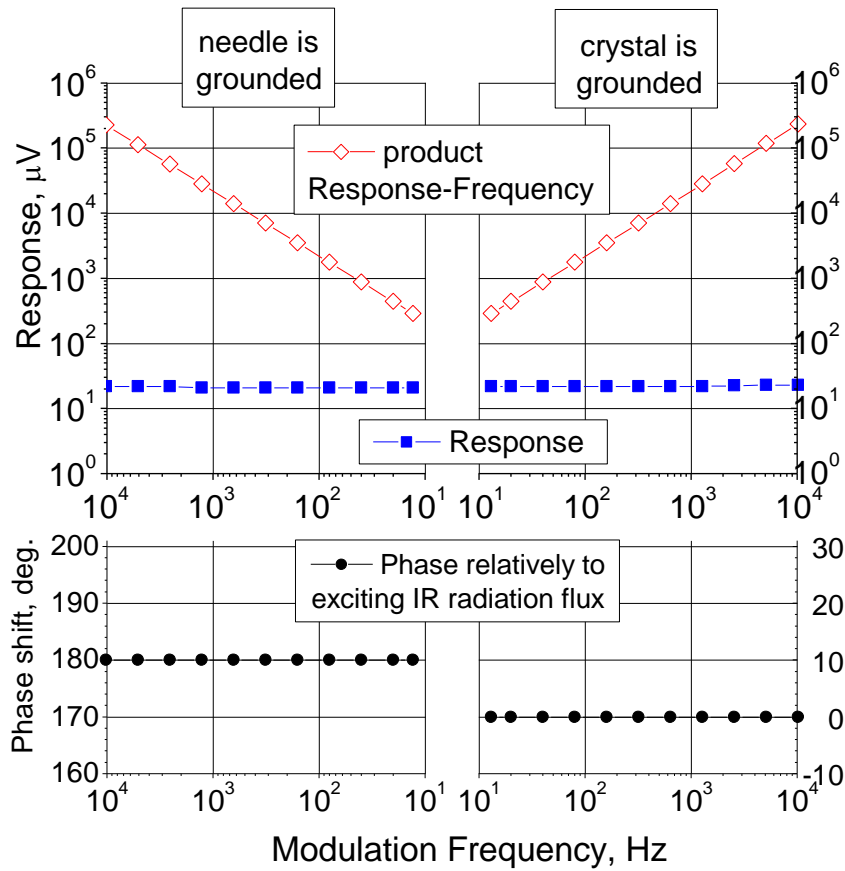

"D1A" high frequency Ge diode
in transparent glass casing:
C = 1,5 pF;
Open circuit electric reponse on modulated
IR-radiation flux (950 nm, 3 mW/mm$^2$)
and response-flux phase shift
(measurements without bias voltage)

For open circuit pyroelectric response of uniformly poled sample
must be $U_\pi(f_m) \sim 1/f_m$; $U_\pi f_m \sim \gamma/c_1\varepsilon$, $\varphi_\pi(f_m) = (+/-)90° = $ const
($\gamma$ is the pyroelectric coefficient, $c_1$ is the volume heat capacity,
$\varepsilon$ is the dielectric permittivity, $f_m$ is the modulation frequency )

**(b)**

**Figure 8.** Dependences of the open circuit electric response on the modulation frequency of IR-irradiation of both sides of pyroelectric non-uniformly poled polyethylene sheet **(a)** and of high frequency Ge diode of D1A type **(b)**.

## 5. Theoretical Modeling

There are many effective medium approximations (shortly "EMA") [29], among which the most known are the Landau approximation of linear mixture [30], Maxwell-Garnett [31] and Bruggeman [32] approximations for spherical inclusions, and Lichtenecker-Rother approximation of the logarithmic mixture [33]. Most of these approximations are applicable for quasi-spherical randomly distributed dielectric (or semiconducting) particles in the insulating environment. These EMA models lead to analytical expressions for the effective permittivity $\varepsilon^*_{eff}$ [34]. However, their applicability ranges are very sensitive to the cross-interaction effects of the polarized ferroelectric particles, and therefore most of them are invalid for ferroelectric mixtures and/or colloids, where the content of the particles is more than (20–30) vol.%.

EMA models for arbitrary content of the nanoparticles were developed by Petzelt et al. [35] and Rychetský et. al. [36]. As a rule, EMA considers the quadratic equation for the effective permittivity of the binary mixture:

$$(1-\mu)\frac{\varepsilon^*_{eff}-\varepsilon^*_b}{(1-n_a)\varepsilon^*_{eff}+n_a\,\varepsilon^*_b} + \mu\frac{\varepsilon^*_{eff}-\varepsilon^*_a}{(1-n_a)\varepsilon^*_{eff}+n_a\,\varepsilon^*_a} = 0. \tag{1}$$

Here $\varepsilon^*_a$, $\varepsilon^*_b$ are relative complex permittivity of the components "*a*" and "*b*" respectively, $\mu$ and $1-\mu$ are relative volume fractions of the components "*a*" and "*b*" respectively, and $n_a$ is the depolarization field factor for the inclusion of the type "*a*".

Following EMA, we assume that BaTiO$_3$ nanoparticles and P(VDF-TrFE) matrix have a complex dielectric permittivity:

$$\varepsilon^*_a(T,\omega) = \varepsilon_a(T,\omega) - i\frac{\sigma_a(T)}{\omega}, \qquad \varepsilon^*_b(T,\omega) = \varepsilon_b(T,\omega) - i\frac{\sigma_b(T)}{\omega}. \tag{2}$$

In the case of pure P(VDF-TrFE) ($\mu=0$), the solution of Eq.(1) is $\varepsilon^*_{eff} = \varepsilon_{PVDF}$, which we assume to have the form:

$$\varepsilon_{PVDF} = \frac{C_{PVDF}}{\sqrt{(T-T_{PVDF}(\omega))^2 + \Delta^2_{PVDF}(\omega)}} + \varepsilon^0_{PVDF}. \qquad (3)$$

The real part of BaTiO$_3$ nanoparticles dielectric permittivity has the same form as for the P(VDF-TrFE):

$$\varepsilon_{BTO} = \frac{C_{BTO}}{\sqrt{(T-T_{BTO}(\omega))^2 + \Delta^2_{BTO}(\omega)}} + \varepsilon^0_{BTO} \qquad (4)$$

Also, to simplify the calculations, we assumed that the conductivities of the P(VDF-TrFE) film and BaTiO$_3$ nanoparticles are temperature independent. The model is incomplete for quantitative description of the temperature dependence of the dielectric permittivity and requires improvement.

Results of theoretical modeling based on Eqs.(1)-(4) qualitatively agree with the experimental results for P(VDF-TrFE)-BaTiO$_3$ films and give the content of nanoparticles equal to 20±5 vol.% for the sample 2 and 35±5 vol.% for the sample 3. The numbers are in a reasonable agreement with SEM results, but are significantly lower than the nominal content of the nanoparticles in the samples 2-3.

## 6. Prediction of the Negative Capacitance State

The stabilization of ferroelectric films, such as lead zirconate-titanate (Pb$_x$Zr$_{1-x}$TiO$_3$), in the state of negative capacitance (NC) [37] was revealed experimentally more than ten years ago [38]. In particular Khan et. al. demonstrated that the total capacitance of a double-layer capacitor, made of paraelectric strontium titanate (SrTiO$_3$) and ferroelectric Pb$_x$Zr$_{1-x}$TiO$_3$, is greater than it would be for a single-layer capacitor comprising SrTiO$_3$ of the same thickness as used in a double-layer capacitor. Using the ferroelectric NC insulator of an appropriate thickness in the gate stack of a Field-Effect Transistor (FET) has several advantages for its architecture, because the NC insulator requires less energy, which significantly reduces heating of nano-chips with a high density of critical electronic elements [39].

A lot of experimental demonstrations of the NC effect in ferroelectric double-layer capacitors are available [40, 41, 42, 43], but only a few propose semi-analytical expressions for the conditions of the NC state appearance and consider the appearance of the domain structure in the ferroelectric layers (see e.g., Refs. [44, 45, 46, 47]). To the best of our knowledge, the NC state in the films with embedded nanoparticles has not been studied yet.

To explain the dielectric permittivity behavior in the P(VDF-TrFE)-BaTiO$_3$ films, the EMA and the finite element modelling (FEM) are used. The results of theoretical modeling are discussed below.

**Figure 9** presents the results of numerical simulation of electric polarization, potential and field distributions in the P(VDF-TrFE)-BaTiO$_3$ films. Insets **(a)-(d)** correspond to the film with a 38.5 vol.% of nanoparticles, and insets **(e)-(h)** correspond to the film with a 70 vol.% of nanoparticles. It is worth noting that in the case of 70 vol.% of nanoparticles, we assumed that the nanoparticles are located so densely that they touch each other and the boundaries of the sample. It is seen from **Figure 9** that 25-nm nanoparticles split into stripe-like domains, which size and period depend strongly on the distance between the particles. Thinner domains correspond to smaller concentrations of nanoparticles due to the weaker cross-interaction and interfacial screening effects. Therefore, higher depolarization field is created by the polarization of the nanoparticles.

Due to the domain structure inside the nanoparticles (as the main reason), as well as due to the differences in the permittivity of the P(VDF-TrFE) and BaTiO$_3$ nanoparticles (as the secondary reason), the electric field concentrates near the surface of the nanoparticles (which is the interfacial effect). The interfacial effects determine electric field distribution, showing much stronger dipole-dipole interactions in the films with higher content of BaTiO$_3$ nanoparticles.

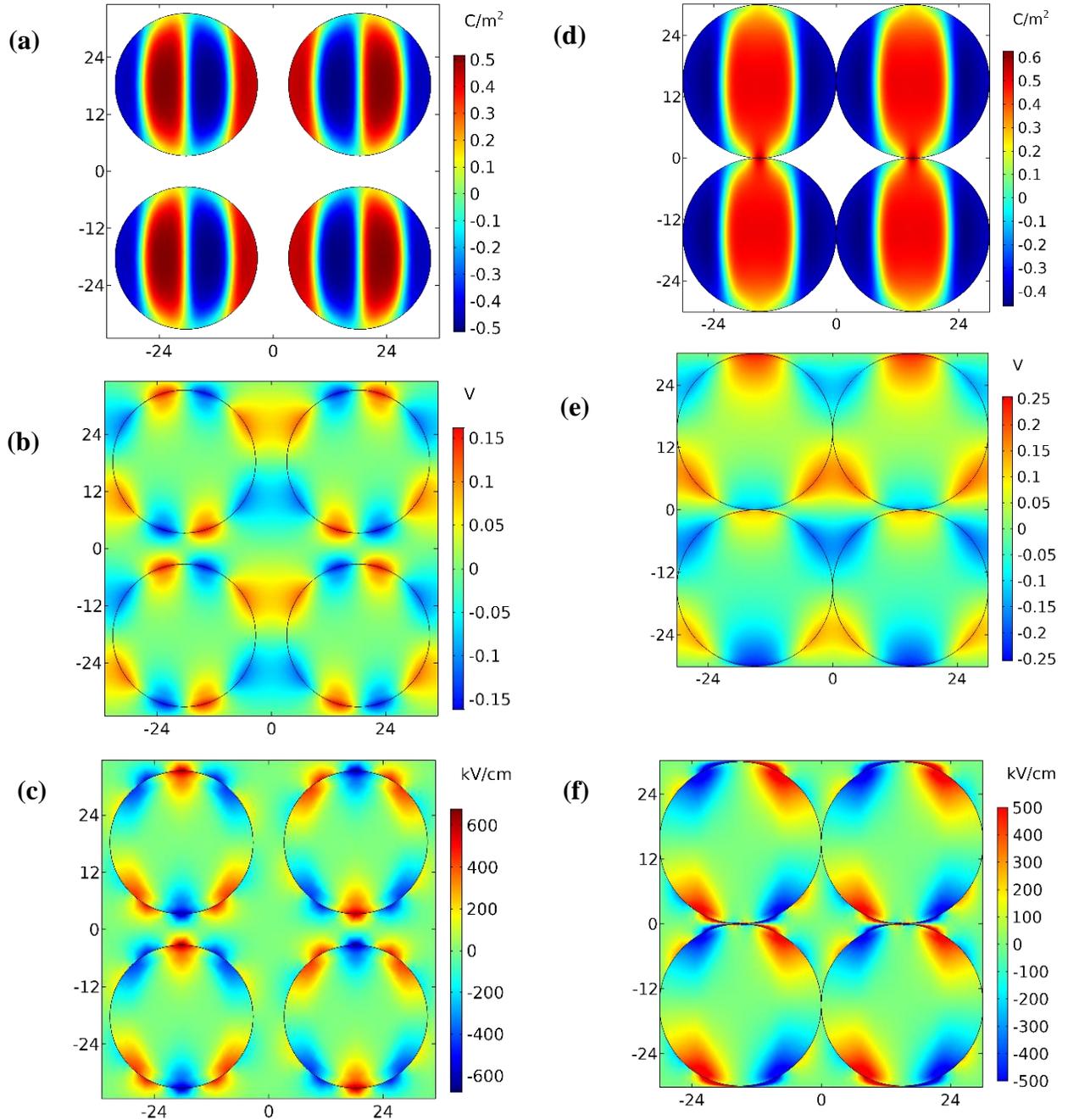

**Figure 9.** FEM of the BaTiO$_3$ nanoparticles polarization, electric potential and field distributions in the middle of the P(VDF-TrFE) film with 38.5 vol. % **(a-d)** and 70

vol. % **(e-h)** of the nanoparticles. The distributions are calculated at zero applied electric field.

Since the average dielectric permittivity of the nanoparticles is greater than that of the matrix, the field is concentrated around them. This leads to the absence of the ferroelectric hysteresis loop in the film with less than 50 vol.% of nanoparticles (not shown here). **Figures 10(a)** and **10(b)** show polarization and dielectric susceptibility versus the applied field (ferroelectric hysteresis) for the P(VDF-TrFE) film with 70 vol.% of nanoparticles. The hysteresis loop for the P(VDF-TrFE) film with the 70 vol.% of nanoparticles, presented in **Fig. 10**, predicts the possibility of the NC state appearance. Indeed, the loop contains two small symmetric loop-type features with a negative slope of polarization curve in the applied field.

We analyzed the conditions for the NC state appearance in the studied P(VDF-TrFE)-$BaTiO_3$ films and obtained that it may appear in a relatively narrow range of the $BaTiO_3$ nanoparticles content (65–75) % and sizes (20–30) nm (at room temperature). Note that the appearance and stability of the NC state also depends strongly on the matrix permittivity and temperature.

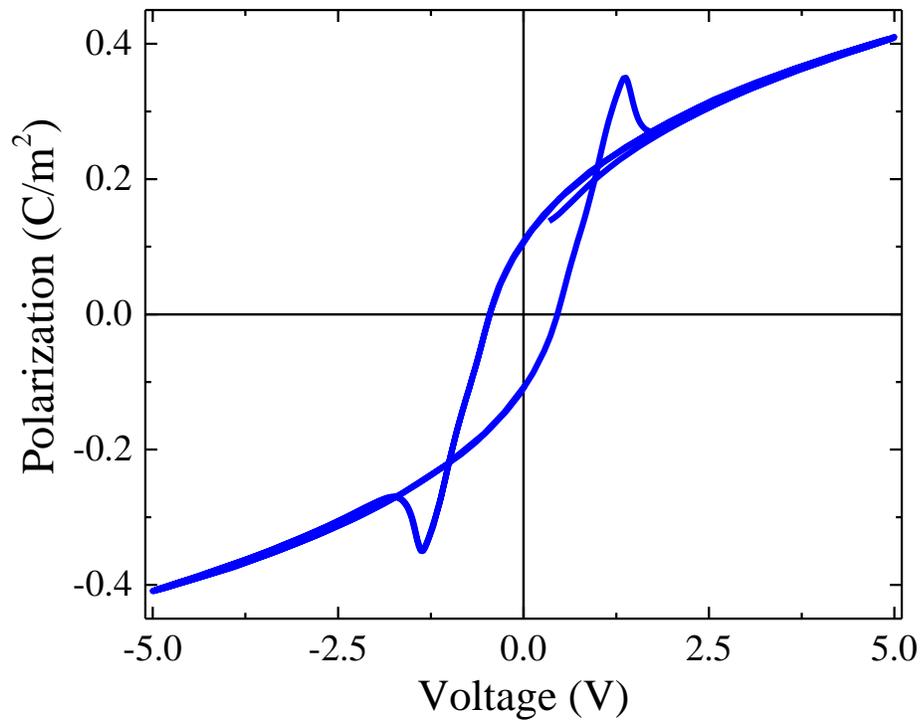

(a)

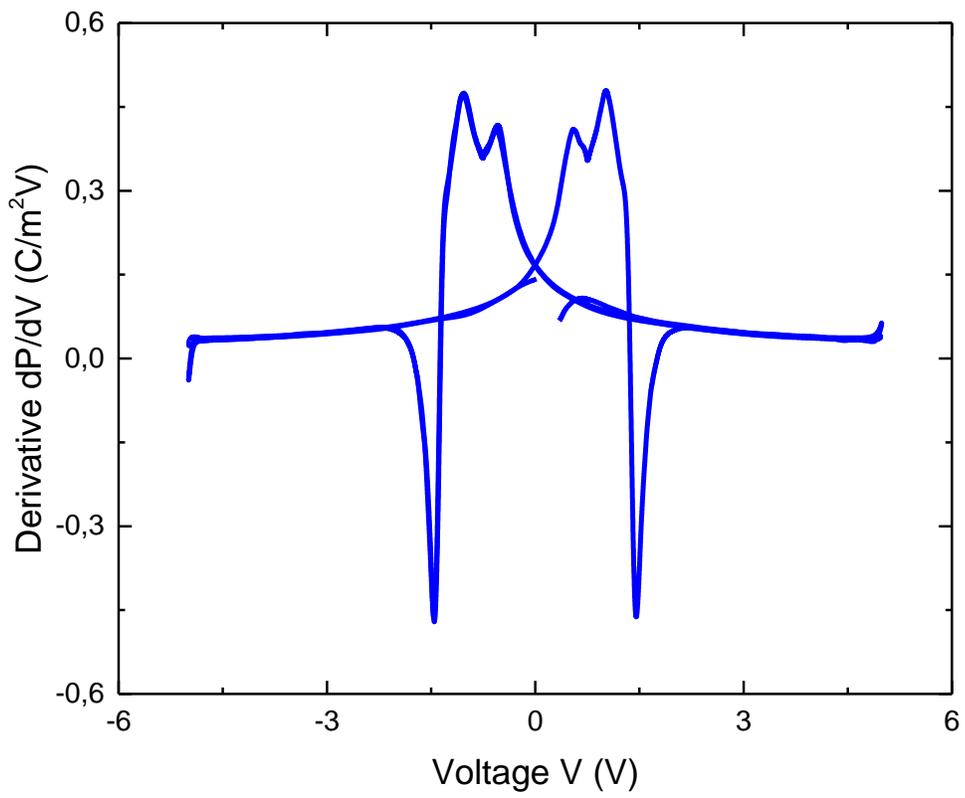

(b)

**Figure 10. (a)** Polarization-field curves in the P(VDF-TrFE) film with 70 vol.% of BaTiO$_3$ nanoparticles. Electric potential was applied to the bottom electrode with a frequency of 1 kHz. **(b)** The derivation dP/dV showing the NC region.

## 7. Conclusions

We analyze the temperature dependences of effective dielectric permittivity of the P(VDF-TrFE) films and layers with BaTiO$_3$ nanoparticles (with the average size 24 nm and dispersion 10 nm). For the P(VDF-TrFE) films with a lower content of BaTiO$_3$ nanoparticles, a diffuse maximum of the temperature dependences of the dielectric permittivity with a relatively strong frequency dispersion is observed in the temperature range (0 – 50)°C. For the P(VDF-TrFE) films with a high content of BaTiO$_3$ nanoparticles, a quasi-plateau of the dielectric permittivity with a relatively weak frequency dispersion is observed in the same temperature range. With increasing the temperature, the dielectric permittivity of P(VDF-TrFE)-BaTiO$_3$ films reveals the following changes: the weak growth at low temperatures, the diffuse step-like change, the diffuse maximum (for the lower content of BaTiO$_3$ nanoparticles) or quasi-plateau (for the higher content of BaTiO$_3$ nanoparticles), and further growth with the temperature increase. The frequency dispersion of the permittivity is relatively weak at low temperatures, moderate in the regions of diffuse step-like changes, and relatively high in the regions of diffuse maxima and in the regions of high-temperature growth.

The content of the ferroelectric nanoparticles extracted from the dielectric response appears significantly different for the nominal filling factors (38 and 70 vol.% of BaTiO$_3$ nanoparticles), indicating on some sort of threshold in the dipole-dipole cross-interaction effects. At the same time, the voltage response of the studied P(VDF-TrFE) - BaTiO$_3$ films to the frequency-modulated IR radiation flux has rather photoelectric than pyroelectric nature.

To explain the dielectric permittivity behavior, the effective medium approximation and the finite element modelling were used. Results of theoretical modeling demonstrated satisfactory agreement with experimental data. Also, we predicted the conditions for the NC state observation in the studied P(VDF-TrFE)-BaTiO$_3$ films and obtained that it may appear in a relatively narrow range of the BaTiO$_3$ nanoparticles volume fraction (65–75) % and sizes (20–30) nm at room temperature.

**Authors' contribution.** The research idea belongs to O.S.P., A.N.M and N.V.M. V.I.S. characterizes the film microstructure by SEM. S.E.I. prepared the P(VDF-TrFE) - BaTiO$_3$ films. D.O.S. prepared the samples for dielectric measurements. O.V.B., D.O.S. and O.S.P. performed dielectric measurements and analyzed obtained results. O.V.B. and E.A.E. wrote the codes and performed FEM. O.V.B. and A.N.M. evolved the analytical model and compared results to experiments. O.V.B., A.N.M. and N.V.M. wrote the manuscript draft. All co-authors discussed the results.

**Acknowledgements.** The work of O.S.P., E.A.E. and A.N.M. is funded by the National Research Foundation of Ukraine (project "Multiple degenerate metastable states of spontaneous polarization in nanoferroics: theory, experiment and prospects for digital nanoelectronics", grant No. 2023.03/0132, and the project "Silicon-compatible ferroelectric nanocomposites for electronics and sensors", grant N 2023.03/0127). The characterization and preparation of P(VDF-TrFE)-BaTiO$_3$ films (S.E.I.), as well as numerical modelling (A.N.M.) is sponsored by the NATO Science for Peace and Security Program under grant SPS G5980 "FRAPCOM". Pyroelectric and the part of electrophysical measurements are sponsored by the Target Program of the National Academy of Sciences of Ukraine, Project No. 5.8/25-П "Energy-saving and environmentally friendly nanoscale ferroics for the development of sensorics, nanoelectronics and spintronics" (O.V.B., D.O.S. and N.V.M.).

# Appendix A

In this work, we neglect the dielectric loss tangent because its value is quite small for the frequences above 1 kHz (see **Fig. A1-A2**). For pure P(VDF-TrFE), the dielectric loss tangent increases from 0.05 to 0.2 for as-prepared samples and decreases for the aged samples. For P(VDF-TrFE) - BaTiO$_3$ films the dielectric loss tangent does not rise above 0.03 - 0.05 for the frequences above 1 kHz.

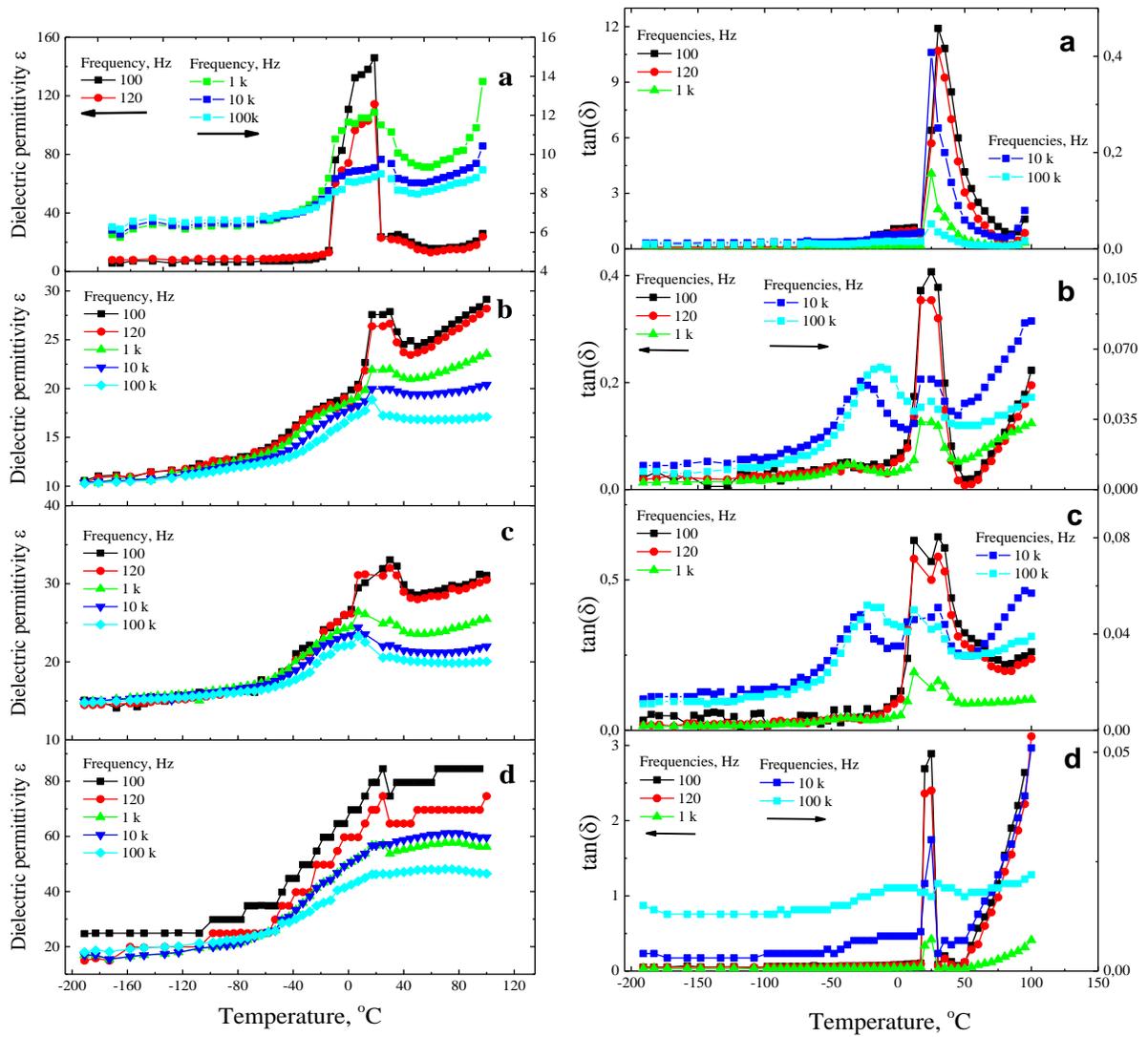

**Figure A1. As prepared samples.** Results of experimental measurements of the temperature dependences of dielectric permittivity and losses for different frequencies for pure P(VDF-TrFE) film of thickness 120 μm **(a)**, P(VDF-TrFE) - BaTiO$_3$ films of thickness 3 μm **(b)**, 8 μm **(c)** and 390 μm **(d)**, respectively.

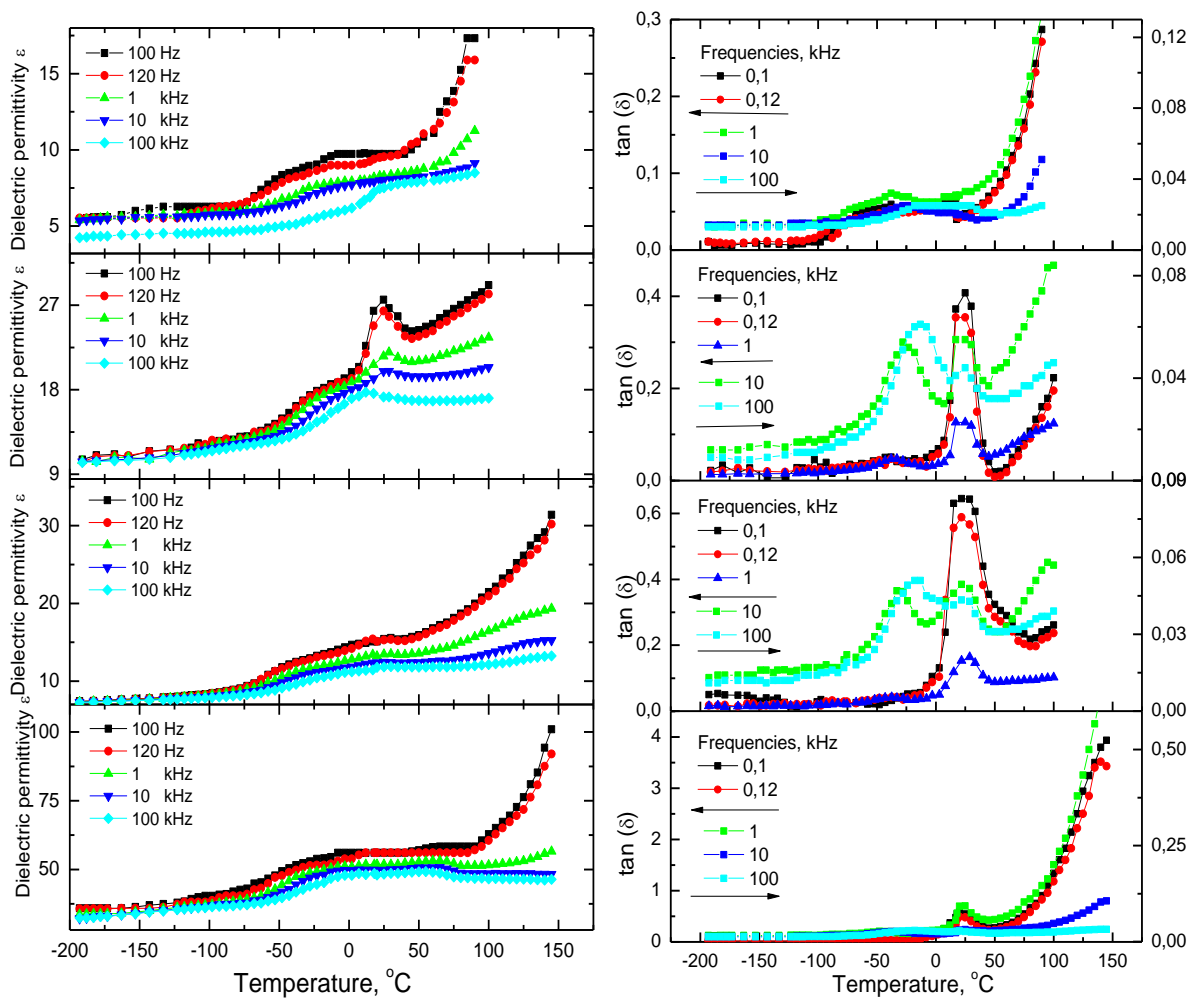

**Figure A2. The samples after six months.** Results of experimental measurements of the temperature dependences of dielectric permittivity and losses for different frequencies for pure P(VDF-TrFE) film of thickness 120 μm **(a)**, P(VDF-TrFE) - BaTiO$_3$ films of thickness 3 μm **(b)**, 8 μm **(c)** and 390 μm **(d)**, respectively.

# References

[1] G. Cook, J. L. Barnes, S. A. Basun, D. R. Evans, R. F. Ziolo, A. Ponce, V. Yu. Reshetnyak, A. Glushchenko, and P. P. Banerjee, Harvesting single ferroelectric domain stressed nanoparticles for optical and ferroic applications. J. Appl. Phys. **108**, 064309 (2010); https://doi.org/10.1063/1.3477163.

[2] D. R. Evans, S. A. Basun, G. Cook, I. P. Pinkevych, and V. Yu. Reshetnyak. Electric field interactions and aggregation dynamics of ferroelectric nanoparticles in isotropic fluid suspensions. Phys. Rev. B, **84,** 174111 (2011); https://doi.org/10.1103/PhysRevB.84.174111.

[3] D. Caruntu and T. Rostamzadeh and T. Costanzo and S. S. Parizi and G. Caruntu. Solvothermal synthesis and controlled self-assembly of monodisperse titanium-based perovskite colloidal nanocrystals, Nanoscale **7**, 12955 (2015).

[4] O. S. Pylypchuk, S. E. Ivanchenko, Y. O. Zagorodniy, M. E. Yelisieiev, O. V. Shyrokov, O. V. Leschenko, O. Bereznykov, D. Stetsenko, S. D. Skapin, E. A. Eliseev, V. N. Poroshin, V. V. Vainberg, and A. N. Morozovska. Relaxor-like Behavior of the Dielectric Response of Dense Ferroelectric Composites. Ceramics Internation (2024), https://doi.org/10.1016/j.ceramint.2024.08.385

[5] Z. Hanani, D. Mezzane, M. Amjoud, M. Lahcini, M. Spreitzer, D. Vengust, A. Jamali, M. El Marssi, Z. Kutnjak, and M. Gouné. "The paradigm of the filler's dielectric permittivity and aspect ratio in high-k polymer nanocomposites for energy storage applications." Journal of Materials Chemistry **C 10**, 10823 (2022); https://doi.org/10.1039/d2tc00251e.

[6] Z. Luo, Z. Ye, B. Duan, G. Li, K. Li, Z. Yang, S. Nie, T. Chen, L. Zhou, and P. Zhai. "SiC@ BaTiO3 core-shell fillers improved high temperature energy storage density of P (VDF-HFP) based nanocomposites." Composites Science and Technology **229**, 109658 (2022); https://doi.org/10.1016/j.compscitech.2022.109658.

[7]    Z. Fan, S. Gao, Y. Chang, D. Wang, X. Zhang, H. Huang, Y. He, and Q. Zhang. "Ultra-superior high-temperature energy storage properties in polymer nanocomposites via rational design of core–shell structured inorganic antiferroelectric fillers." Journal of Materials Chemistry A **11**, 7227 (2023); https://doi.org/10.1039/D2TA09658G.

[8]    G. Zhang, X. Zhang, T. Yang, Qi Li, L.-Q. Chen, S. Jiang, and Q. Wang. Colossal room-temperature electrocaloric effect in ferroelectric polymer nanocomposites using nanostructured barium strontium titanates. ACS nano **9**, 7164 (2015); https://doi.org/10.1021/acsnano.5b03371.

[9]    A. N. Morozovska, O. S. Pylypchuk, S. Ivanchenko, E. A. Eliseev, H. V. Shevliakova, L. M. Korolevich, L. P. Yurchenko, O. V. Shyrokov, N. V. Morozovsky, V. N. Poroshin, Z. Kutnjak, and V. V. Vainberg. Size-induced High Electrocaloric Response of the Dense Ferroelectric Nanocomposites. Ceramics International **50** (7b), 11743 (2024), https://doi.org/10.1016/j.ceramint.2024.01.079

[10]   J. Zhu, W. Han, H. Zhang, Z. Yuan, X. Wang, L. Li, and C. Jin. Phase coexistence evolution of nano $BaTiO_3$ as function of particle sizes and temperatures. J. Appl. Phys. **112**, 064110 (2012); https://doi.org/10.1063/1.4751332.

[11]   U. Idehenre, Y. A. Barnakov, S. A. Basun, and D. R. Evans. Spectroscopic studies of the effects of mechanochemical synthesis on BaTiO3 nanocolloids prepared using high-energy ball-milling. J. Appl. Phys. **124**, 165501 (2018).

[12]   U. Idehenre, Y. A. Barnakov, S. A. Basun, and D. R. Evans. Spectroscopic studies of the effects of mechanochemical synthesis on BaTiO3 nanocolloids prepared using high-energy ball-milling. J. Appl. Phys. **124**, 165501 (2018).

[13]   Y. A. Barnakov, I. U. Idehenre, S. A. Basun, T. A. Tyson, and D. R. Evans. Uncovering the mystery of ferroelectricity in zero dimensional nanoparticles. Nanoscale Advances **1**, 664 (2019).

[14] H. Zhang, S. Liu, S. Ghose, B. Ravel, I. U. Idehenre, Y. A. Barnakov, S. A. Basun, D. R. Evans, and T. A. Tyson. Structural Origin of Recovered Ferroelectricity in BaTiO$_3$ Nanoparticles. Phys. Rev. B **108**, 064106 (2023); https://doi.org/10.1103/PhysRevB.108.064106.

[15] A. N. Morozovska, E. A. Eliseev, S. V. Kalinin, and D. R. Evans. Strain-Induced Polarization Enhancement in BaTiO$_3$ Core-Shell Nanoparticles. Phys.Rev. B. **109**, 014104 (2024), https://link.aps.org/doi/10.1103/PhysRevB.109.014104

[16] J. M. Gudenko, O. S. Pylypchuk, V. V. Vainberg, I. A. Gvozdovskyy, S. E. Ivanchenko, D. O. Stetsenko, N. V. Morozovsky, V. N. Poroshin, E. A. Eliseev and A. N. Morozovska, Ferroelectric Nanoparticles in Liquid Crystals: The Role of Ionic Transport at Small Concentrations of the Nanoparticles. Semiconductor Physics, Optoelectronics and Quantum Electronics **28**, N 1, 010-018 (2025), https://doi.org/10.15407/spqeo28.01.010

[17] M. E. Lines and A. M. Glass, Principles and Application of Ferroelectrics and Related Materials (Clarendon Press, Oxford, 1977).

[18] O. S. Pylypchuk, S. E. Ivanchenko, M. Y. Yelisieiev, A. S. Nikolenko, V. I. Styopkin, B. Pokhylko, V. Kushnir, D. O. Stetsenko, O. Bereznykov, O. V. Leschenko, E. A. Eliseev, V. N. Poroshin, N. V. Morozovsky, V. V. Vainberg, and A. N. Morozovska. Behavior of the Dielectric and Pyroelectric Responses of Ferroelectric Fine-Grained Ceramics. Journal of the American Ceramic Society (2025), https://doi.org/10.1111/jace.20391

[19] H. Arisawa , O. Yano & Y. Wada, Dielectric loss of poly(vinylidene fluoride) at low temperatures and effect of poling on the low temperature loss. Ferroelectrics, **32** (1), 39 (1981); https://doi.org/10.1080/00150198108238671

[20] Y. Takase, H. Tanaka, T. T. Wang, R. E. Cais, and J. M. Kometani, Ferroelectric properties of form 1 perdeuteriated poly(vinylidene fluoride), Macromolecules, **20**, 2318 (1987); https://doi.org/10.1021/ma00175a049
38

[21] B. Hilczer, J. Kulek, E. Markiewicz, M. Kosec and B. Malic, Dielectric relaxation in ferroelectric PZT-PVDF nanocomposites. J. Non-Cryst. Solids, **305**, 167 (2002); https://doi.org/10.1016/S0022-3093(02)01103-1

[22] V. A. Stephanovich, M. D. Glinchuk, E. V. Kirichenko, B. Hilczer, Theory of radiation induced relaxor behavior of poly.vinylidene fluoride-trifluoroethylene. copolymers. J. Appl. Phys., **94** (9), 5937 (2004); https://doi.org/10.1063/1.1613810

[23] T. Yagi, M. Tatemoto, and J. Sako, Transition behavior and dielectric properties in trifluoroethylene and vinylidene fluoride copolymers. Polymer Journal, **12** (4), 209 (1980); https://doi.org/10.1295/polymj.12.209

[24] V. S. Yadav, D. K. Sahu, Y. Singh, and D. C. Dhubkarya, "The effect of frequency and temperature on dielectric properties of pure poly Vinylidene tluoride (P(VDF-TrFE)) Thin Films", Proceedings of International MultiConference of Engineers of Computer Scientists 2010 Vol.III, IMECS 2010, March 17-19, 2010, Hong Kong. ISBN: 978-988-18210-5-8; ISSN: 2078-0958 (Print); ISSN: 2078-0966 (Online)

[25] Q. Jiang, X. F. Cui, M. Zhao, "Size effects on Curie temperature of ferroelectric particles", Appl. Phys. A 78, 703-704 (2004); https://doi.org/10.1007/s00339-002-1959-6

[26] F. Jona and G. Shirane, Feroelectric Crystals (Macmillan, New York, 1962).

[27] S. L. Bravina, N. V. Morozovsky, A. A. Strokach, "Pyroelectricity: some new research and application aspects." In Material Science and Material Properties for Infrared Optoelectronics, F. F. Sizov (Ed.), v. **3182**, pp. 85-99, SPIE: Bellingham, (1996), https://doi.org/10.1117/12.280409

[28] L. S. Kremenchugsky, *Ferroelectric Detectors of Radiation*, Naukova Dumka, Kyiv, 1971.

[29] T. C. Choy, Effective Medium Theory. Oxford: Clarendon Press. 1999. ISBN 978-0-19-851892-1.

[30]     L. D. Landau, L. P. Pitaevskii, E. M. Lifshitz Electrodynamics of continuous media. 2013. **8**. P. 475. Elsevier. (Translated by J. S. Bell, M. J. Kearsley, and J. B. Sykes).

[31]     J. C. M. Garnett, "Colours in metal glasses and in metallic films," Philos. Trans. R. Soc. London. Ser. A **203**, 385. 1904. https://doi.org/10.1098/rsta.1904.0024.

[32]     D. A. G. Bruggeman, "Berechnung verschiedener physikalischer Konstanten von heterogenen Substanzen. I. Dielektrizitätskonstanten und Leitfähigkeiten der Mischkörper aus isotropen Substanzen," Ann. Phys. 1935. **416**. P. 636. https://doi.org/10.1002/andp.19354160705.

[33]     R. Simpkin, Derivation of Lichtenecker's logarithmic mixture formula from Maxwell's equations. IEEE Transactions on Microwave Theory and Techniques. 2010. **58.** P. 545. https://doi.org/10.1109/TMTT.2010.2040406.

[34]     G. L. Carr, S. Perkowitz, and D. B. Tanner. "Far-infrared properties of inhomogeneous materials."in: Infrared and millimeter waves. 13, ed. K.J. Button, Academic Press, Orlando. (1985): 171-263.

[35]     J. Petzelt, D. Nuzhnyy, V. Bovtun, D.A. Crandles. Origin of the colossal permittivity of (Nb+ In) co-doped rutile ceramics by wide-range dielectric spectroscopy. Phase Transitions. 2018. **91.** P. 932. https://doi.org/10.1080/01411594.2018.1501801.

[36]     I. Rychetský, D. Nuzhnyy, J. Petzelt, Giant permittivity effects from the core–shell structure modeling of the dielectric spectra. Ferroelectrics. 2020. **9.** P. 569. https://doi.org/10.1080/00150193.2020.1791659.

[37]     A. I. Khan, K. Chatterjee, B. Wang, S. Drapcho, L. You, C. Serrao, S. R. Bakaul, R. Ramesh, and S. Salahuddin. Negative capacitance in a ferroelectric capacitor. Nat. Mater. **14**, 182 (2015), https://doi.org/10.1038/nmat4148

[38]     A.I. Khan, D. Bhowmik, P. Yu, S. J. Kim, X. Pan, R. Ramesh, and S. Salahuddin. Experimental evidence of ferroelectric negative capacitance in nanoscale

heterostructures. Appl. Phys. Lett. **99**, 113501 (2011), https://doi.org/10.1063/1.3634072

[39] IRDS™ 2023: Beyond CMOS, https://irds.ieee.org/editions/2023

[40] P. Zubko, J. C. Wojdeł, M. Hadjimichael, S. Fernandez-Pena, A. Sené, I. Luk'yanchuk, J.-M. Triscone, and J. Íñiguez. Negative capacitance in multidomain ferroelectric superlattices. Nature **534**, 524 (2016), https://doi.org/10.1038/nature17659

[41] M. Hoffmann, M. Pešic, S. Slesazeck, U. Schroeder, and T. Mikolajick, On the stabilization of ferroelectric negative capacitance in nanoscale devices. Nanoscale **10**, 10891 (2018), https://doi.org/10.1039/C8NR02752H

[42] A.K. Yadav, K. X. Nguyen, Z. Hong, P. García-Fernández, P. Aguado-Puente, C. T. Nelson, S. Das, B. Prasad, D. Kwon, S. Cheema, A. I. Khan, C. Hu, Jorge Íñiguez, J. Junquera, L.-Q. Chen, D. A. Muller, R. Ramesh, S. Salahuddin. Spatially resolved steady-state negative capacitance. Nature **565**, 468 (2019), https://doi.org/10.1038/s41586-018-0855-y

[43] S.M. Neumayer, L. Tao, A. O'Hara, M.A. Susner, M.A. McGuire, P. Maksymovych, S.T. Pantelides, and N. Balke. The concept of negative capacitance in ionically conductive van der Waals ferroelectrics. Adv. Energy Mater. **10**, 2001726 (2020), https://doi.org/10.1002/aenm.202001726

[44] E. A. Eliseev, M. E. Yelisieiev, S. V. Kalinin, and Anna N. Morozovska. Observability of negative capacitance of a ferroelectric film: Theoretical predictions. Phys. Rev. B **105**, 174110 (2022), https://doi.org/10.1103/PhysRevB.105.174110

[45] I. Luk'yanchuk, A. Razumnaya, A. Sene, Y. Tikhonov, and V. M. Vinokur. The ferroelectric field-effect transistor with negative capacitance. npj Computational Materials **8**, 52 (2022), https://doi.org/10.1038/s41524-022-00738-2

[46] I.B. Misirlioglu, M. K. Yapici, K. Sendur, and M. B. Okatan. Weak Dependence of Voltage Amplification in a Semiconductor Channel on Strain State and Thickness of

a Multidomain Ferroelectric in a Bilayer Gate. ACS Applied Electronic Materials **5**, 6832 (2023), https://doi.org/10.1021/acsaelm.3c01271

[47] A. N. Morozovska, E. A. Eliseev, O. A. Kovalenko, and D. R. Evans. The Influence of Chemical Strains on the Electrocaloric Response, Polarization Morphology, Tetragonality and Negative Capacitance Effect of Ferroelectric Core-Shell Nanorods and Nanowires. Physical Review Applied **21**, 054035 (2024), https://doi.org/10.1103/PhysRevApplied.21.054035
42